\documentclass[letterpaper]{article}
\usepackage{aaai25} % DO NOT CHANGE THIS
\usepackage{times} % DO NOT CHANGE THIS
\usepackage{helvet} % DO NOT CHANGE THIS
\usepackage{courier} % DO NOT CHANGE THIS
\usepackage[hyphens]{url} % DO NOT CHANGE THIS
\usepackage{graphicx} % DO NOT CHANGE THIS
\usepackage{graphicx} % DO NOT CHANGE THIS
\usepackage{natbib} % DO NOT CHANGE THIS
\usepackage{caption} % DO NOT CHANGE THIS
\usepackage{graphicx}
\usepackage{tikz}
\usepackage{subfigure}
\usepackage{pgf}
\usepackage{adjustbox}
\usepackage{todonotes}

\usetikzlibrary{arrows,
	automata,
	calc,
	decorations,
	decorations.pathreplacing,
	positioning,
	shapes,
	calligraphy}
\tikzset{res/.style={ellipse,draw,minimum height=0.5cm,minimum width=0.8cm}}

\usepackage{amsthm}

\newcommand{\Ende}{\hfill $\lhd$}
\usepackage{thmtools}

\usepackage{amsmath}
\usepackage{amsfonts}
\usepackage{amssymb}
\usepackage{eulervm}

\newtheoremstyle{italic}
{9pt}%      Space above, empty = `usual value'
{9pt}%      Space below
{\itshape}% Body font
{}%         Indent amount (empty = no indent, \parindent = para indent)
{\bfseries}% Thm head font
{.}%        Punctuation after thm head
{.5em}% Space after thm head: \newline = linebreak
{\bfseries{\thmname{#1}\thmnumber{ #2}\thmnote{ (#3)}}}

\newtheoremstyle{normal}
{9pt}%      Space above, empty = `usual value'
{9pt}%      Space below
{}% Body font
{}%         Indent amount (empty = no indent, \parindent = para indent)
{\bfseries}% Thm head font
{.}%        Punctuation after thm head
{.5em}% Space after thm head: \newline = linebreak
{\bfseries{\thmname{#1}\thmnumber{ #2}\thmnote{ (#3)}}}

\theoremstyle{normal}
\newtheorem{Def}{Definition}

\newtheorem{definition}[Def]{Definition}

\newtheorem{remark}[Def]{Remark}
\newtheorem{example}[Def]{Example}

\theoremstyle{italic}

\newtheorem{lemma}[Def]{Lemma}
\newtheorem{theorem}[Def]{Theorem}
\newtheorem{proposition}[Def]{Proposition}
\newtheorem{corollary}[Def]{Corollary}

\newtheorem*{claim*}{Claim}

\newcommand{\condG}{\textbf{(G)} }
\newcommand{\condS}{\textbf{(S)} }
\newcommand{\condR}{\textbf{(R)} }

\newcommand{\Cause}{\mathit{C}}

\newcommand{\Effect}{\mathit{E}}
\newcommand{\Eff}{\mathit{E}}

\newcommand{\eff}{\mathit{eff}}

\newcommand{\tp}{\mathit{tp}}
\newcommand{\fp}{\mathit{fp}}
\newcommand{\fn}{\mathit{fn}}
\newcommand{\tn}{\mathit{tn}}
\newcommand{\TP}{\mathit{TP}}
\newcommand{\FP}{\mathit{FP}}
\newcommand{\FN}{\mathit{FN}}
\newcommand{\TN}{\mathit{TN}}

\newcommand{\precision}{\operatorname{\mathit{precision}}}
\newcommand{\recall}{\operatorname{\mathit{recall}}}
\newcommand{\fscore}{\operatorname{\mathit{fscore}}}

\newcommand{\mcc}{\operatorname{mcc}}

\newcommand{\canMDP}[2]{#1_{[#2]}}
\newcommand{\minmaxMDP}[2]{#1_{<#2>}}
\newcommand{\twoMDP}[2]{#1_{#2}}

\newcommand{\freq}[2]{\mathit{freq}_{#1}(#2)}

\newcommand{\Nat}{\mathbb{N}}
\newcommand{\Real}{\mathbb{R}}

\def\Pr{\mathrm{Pr}}

\newcommand{\eqdef}{=}

\newcommand{\init}{\mathit{init}}

\newcommand{\StAct}{\mathit{StAct}}
\newcommand{\SA}{\StAct}

\newcommand{\Act}{\mathit{Act}}

\newcommand{\residual}[2]{\mathit{res}(#1,#2)}

\newcommand{\sched}{\mathfrak{S}}

\newcommand{\tsched}{\mathfrak{T}}
\newcommand{\usched}{\mathfrak{U}}

\newcommand{\Paths}{\mathit{Paths}}
\newcommand{\fin}{\mathit{fin}}

\newcommand{\last}{\mathit{last}}

\newcommand{\Until}{ \, \mathrm{U}\, }
\newcommand{\until}{\Until}

\newcommand{\PTIME}{\mathrm{P}}

\newcommand{\NP}{\mathrm{NP}}

\def\cE{\mathcal{E}}

\def\cM{\mathcal{M}}
\def\cN{\mathcal{N}}

\usepackage{thmtools}

\begin{document}
\title{Formal Quality Measures for Predictors in Markov Decision Processes}
\author{
Chistel Baier\textsuperscript{\rm 1},
Sascha Kl\"uppelholz\textsuperscript{\rm 1}
Jakob Piribauer\textsuperscript{\rm 1,2},
Robin Ziemek\textsuperscript{\rm 1}
}
\affiliations{
\textsuperscript{\rm 1}Technische Universit\"at Dresden,\\
\textsuperscript{\rm 2}Universit\"at Leipzig,\\
\{christel.baier, sascha.klueppelholz, jakob.piribauer, robin.ziemek\}@tu-dresden.de
}

\maketitle
 \begin{abstract}
 	In adaptive systems, predictors are used to anticipate changes in the system's state or behavior that may require system adaption, e.g., changing its configuration or adjusting resource allocation.
 	Therefore, the quality of predictors is crucial for the overall reliability and performance of the system under control. This paper studies predictors in systems exhibiting probabilistic and non-deterministic behavior modelled as Markov decision processes (MDPs).
 	Main contributions  are the introduction of quantitative notions that measure the effectiveness of predictors in terms of their average capability to predict the occurrence of failures or other undesired system behaviors.  	The average is taken over all memoryless policies.
 	We study two classes of such notions.
 	One class is inspired by concepts that have been introduced in statistical analysis to explain the impact of features on the decisions of binary classifiers (such as precision, recall, f-score).
 	Second, we study a measure that borrows ideas from recent work on probability-raising causality in MDPs and determines the quality of a predictor by the fraction of memoryless policies under which (the set of states in) the predictor is a probability-raising cause for the considered failure scenario.
\end{abstract}

\section{Introduction}

In modern days, AI systems grow ever more complex and harder to understand, e.g. code designed by artificial intelligence tends to be very abstruse and thus is not comprehensible in a simple way.
Since a full understanding of such systems is difficult to establish, it is important to predict certain events within such systems.
In particular, situations in which the system produces unwanted or even disastrous results need to be predicted early and precisely.

In the area of formal verification, counterexamples, invariants and related certificates are often used to provide a verifiable justification that a system does or does not behave according to a specification (see e.g., \cite{MaPn95,CGP99,Namjoshi01}).
However, most AI systems can not be designed in a way that failure can be excluded and then certificates do not provide enough insights on the systems decisions to predict its behavior.
In order to get an understanding \emph{why} a system behaves the way it does, we introduce measures in how well certain events in a system serve as a predictor for undesired outcomes.
Events which have a cause-effect relation to such outcomes constitute a special case of such predictors \cite{HalpernP04, Pearl09}.

In this paper, we consider binary predictors in Markov decision processes (MDPs) which are a stochastic operational model with non-deterministic choices.
We interpret the non-determinism as uncertainty about the future behavior and thus it may or may not be resolved adversarial to our goals.

\begin{figure}[t]
	\centering
	\resizebox{0.35\textwidth}{!}{
		\begin{tikzpicture}[scale=1,->,>=stealth',auto ,node distance=0.5cm, thick]
			\tikzstyle{round}=[thin,draw=black,circle]
			
			\node[scale=1, state] (init) {$\init$};
			\node[scale=1, state, below=0.4 of init] (end) {$\mathit{end}$};
			\node[scale=1, state, left=2 of init] (b) {$\mathit{Billy}$};
			\node[scale=1, state, right=2 of init] (s) {$\mathit{Suzy}$};
			\node[scale=1, ellipse, draw, below=1.88 of init] (h) {$\mathit{Shatter}$};
			\node[scale=1, state, below=1.7 of b] (bt) {$\mathit{BT}$};
			\node[scale=1, state, below=1.7 of s] (st) {$\mathit{ST}$};
			
			\draw[<-] (init) --++(-0.55,0.55);
			
			\draw[color=black ,->] (init) edge node [pos=0.5, right] {$\frac{1}{5}$} (end);
			\draw[color=black ,->] (init) edge node [pos=0.5, above] {$\frac{2}{5}$} (b);
			\draw[color=black ,->] (init) edge node [pos=0.5, above] {$\frac{2}{5}$} (s);
			
			\draw[color=black ,->] (b) edge node [pos=0.5, right] {$t$} (bt);
			\draw[color=black ,->] (b) edge[out=330, in=210] node [pos=0.5, above] {$w$} (init);
			
			\draw[color=black ,->] (s) edge node [pos=0.5, right] {$t$} (st);
			\draw[color=black ,->] (s) edge[out=210, in=330] node [pos=0.5, above] {$w$} (init);
			
			\draw[color=black ,->] (bt) edge node [pos=0.5, above] {$\frac{1}{2}$}(end);
			\draw[color=black ,->] (bt) edge node [pos=0.5, below] {$\frac{1}{2}$} (h);
			
			\draw[color=black ,->] (st) edge node [pos=0.5, above] {$\frac{1}{5}$} (end);
			\draw[color=black ,->] (st) edge node [pos=0.5, below] {$\frac{4}{5}$} (h);
			
		\end{tikzpicture}
	}
	\caption{An experiment where we want to predict $\mathit{Shatter}$}
	\label{fig:phil-actual}
\end{figure}
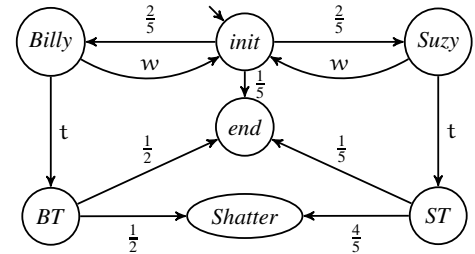

For example, consider an experiment with two participating persons ``Suzy'' and ``Billy'' which are asked to throw a rock at a bottle of glass.
This example has been widely discussed for in philosophic literature on \emph{causality} \cite{Hall2004, ChocklerH04, Halpern15}.
In our variant (Fig. \ref{fig:phil-actual}) a randomized process decides, which person is allowed to take a throw or whether the experiment ends.
However, the participants can choose to wait $w$ or to throw $t$.
If someone decides to throw their rock, the state is changed accordingly ($\mathit{ST}$ for ``Suzy throws''  and $\mathit{BT}$ for ``Billy throws'').
For predictions in this system, we do not make any assumptions on the decisions of the participants and consider both the outcomes and predictors described by sets of states.
For example, reaching the state $\mathit{ST}$ is intuitively a good predictor since the throw of Suzy has a high probability of hitting.
However, if Suzy does not feel confident and thus has a  decides to throw with a low probability, then state $\mathit{ST}$ will only be reached with low probability and so a lot of scenarios in which the bottle shatters come from Billys throw.
In order to distinguish between the quality of different predictors, we use measures from statistical analysis \cite{Powers-fscore}.
However, the quality of a prediction may rely on the distribution over the decisions, as e.g. between $w$ and $t$ for $\mathit{Suzy}$.
So, we consider an \emph{average} case scenario with respect to the non-determinism for the quality of a prediction.

Furthermore, we also consider whether a chosen predictor has a cause-effect relation to the undesired outcome.
Such a probabilistic cause-effect relation in MDPs is introduced in \cite{LMCS24} where the probability-raising (PR) principle is invoked for each possible resolution of the non-determinism.
Inspired by this, we introduce \emph{probability-raising policies}, which witness a PR condition.
A predictor can then also be rated by the relative amount of resolutions (the \emph{causal volume}) in which it has a probabilistic cause-effect relation with the predicted event.

\paragraph{Contributions}
By considering binary predictors in MDPs we formally introduce an average case analysis for quality measures depending on the non-determinism in order to rate the quality of a predictor (Sec. \ref{sec:Quality}).
For this we use a uniform measure over the memoryless randomized policies.
We also introduce the concept of probability-raising policies (Sec. \ref{sec:PR-policy}), which we use to define causal volumes for predictors in MDPs as an additional way to get information about the quality of a predictor (Sec. \ref{sec:causal-volume}).
We then address the complexity of deciding the existence of PR policies (Sec. \ref{sec:PR-check}).

\paragraph{Related Work}
Considering the quality of a prediction has connections to responsibility \cite{ChocklerH04, ChocklerHK2008}, blameworthiness \cite{Halpern2018TowardsFD} and harm \cite{BCH23}.
In \cite{MBFJK21-Importance} and \cite{IJCAI21} forward responsibilities based on the Shapley value are allocated in Kripke and game structures.
Recent work considers backwards responsibility in deterministic AI systems \cite{AAAI24}.

All these notions of responsibility are based on causality as a necessary condition \cite{BrahamvanHees2012}.
A causality based account of responsibility can be found in \cite{ChocklerHK2008}.
In stochastic operational models, probabilistic causes are used as predictors in \cite{ISSE22} and interpreted as binary classifiers in \cite{LMCS24}.

Predicting events in Markovian models also has connections to monitoring properties.
The fact that randomization improves monitors for non-probabilistic systems has been examined in \cite{ChadhaSV09}. 
A current risk-value is estimated for states in partially observable MDPs in \cite{JTS-RuntimeMonitorsMDP2021}.

\section{Preliminaries}
\label{sec:prelim}

In the context of this work a \emph{Markov decision process (MDP)} is a 4-tuple $\cM=(S, \Act,P,\init)$ where $S$ is a finite set of states, $\Act$ a finite set of actions, $\init \in S$ the initial state and $P : S \times \Act \times S \to [0,1]$ the probabilistic transition function such that $\sum_{t\in S}P(s, \alpha,t) \in \{0,1 \}$ for all states $s \in S$ and actions $\alpha \in \Act$.
An action $\alpha$ is \emph{enabled} in state $s \in S$ if $\sum_{t\in S}P(s, \alpha,t)=1$ and $\Act(s)$ denotes the set of enabled actions in $S$.
A state $t$ is \emph{terminal} if $\Act(t) = \emptyset$.
A \emph{path} in an MDP $\cM$ is a (finite or infinite) alternating sequence $\pi=s_0 \, \alpha_0 \, s_1 \, \alpha_1 \, s_2 \dots \in (S \times \Act)^* \cup (S \times \Act)^\omega$ such that $P(s_i, \alpha_{i},s_{i+1})>0$ for all indices $i$.
A path is called maximal if it is infinite or finite and ends in a terminal state.
An MDP can be seen as a Kripke structure in which transitions go from states to probability distributions over states.

A \emph{(randomized) policy} $\sched$ is a function that maps each finite non-maximal path $s_0 \alpha_0  \dots \alpha_{n-1} s_n$ to a distribution over $\Act(s_n)$.
$\sched$ is called deterministic if $\sched(\pi)$ is a Dirac distribution for all finite non-maximal paths $\pi$.
If the chosen action only depends on the last state of the path, $\sched$ is called \emph{memoryless}.
We write MR for the class of memoryless (randomized) and MD for the class of memoryless deterministic policies.
\emph{Finite-memory} policies are those that are representable by a finite-state automaton.

A policy $\sched$ of $\cM$ induces a (possibly infinite) Markov chain.
We write $\Pr^{\sched}_{\cM,s}$ for the standard probability measure on measurable sets of maximal paths in the Markov chain induced by $\sched$ with initial state $s$.
We use the abbreviation $\Pr^{\sched}_{\cM}=\Pr^{\sched}_{\cM,\init}$.
We use linear temporal logic (LTL) modalities such as $\Diamond$ (eventually) and $\Until$ (until) to denote path properties.
For $X,T \subseteq S$ the formula $X \Until T$ is satisfied by $\pi = s_0 s_1 \dots $ if there is $j \geq 0$ such that for all $i<j: s_i \in X$ and $s_j \in T$ and $\Diamond T = S \Until T$.
It is well-known that $\Pr^{\min}_{\cM}(X \Until T)$ and $\Pr^{\max}_{\cM}(X \Until T)$ and corresponding optimal MD-policies are computable in polynomial time.

For $s\in S$ and $\alpha\in \Act(s)$, $(s,\alpha)$ is a state-action pair of $\cM$.
We denote the set of state-action pairs of $\cM$ by $\StAct$.
An \emph{end component} (EC) of an MDP $\cM$ is a strongly connected sub-MDP containing at least one state-action pair.

For a policy $\sched$ of $\cM$, the expected \emph{frequencies} of state-action pairs $(s, \alpha)$ are
\begin{center}
	$\freq{\sched}{s,\alpha} \eqdef \mathrm{E}^{\sched}_{\cM}(\text{no. of visits to $s$ in which $\alpha$ is taken})$
\end{center}
In end-component free MDPs we can specify MR policies by their state-action frequencies (see e.g. \cite[Theorem 4.7]{Kallenberg20}), by considering a linear constraint system over variables $x_{s,\alpha}$ for each $(s, \alpha) \in \StAct$:
\begin{align*}
	\label{eq:balance}
	x_{s, \alpha} & \geq \ 0 \qquad \text{for all $(s, \alpha) \in \SA$},
	\tag{S1}
	\\
	x_{\init} & \ = \
	1+ \!\!\!\!\! \sum_{(t,\alpha) \in \SA} \!\!\!\!\!
	x_{t,\alpha}\cdot P(t,\alpha,\init),
	\tag{S2}\\
	x_{s} & \ = \!\!\!\!\! \sum_{(t,\alpha) \in \SA} \!\!\!\!\! x_{t,\alpha}\cdot P(t,\alpha,s)
	\ \ \ \text{for all $s\in S\setminus\{\init\}$},
	\tag{S3}
\end{align*}
where we use the short form notation $x_s = \sum_{\alpha \in \Act(s)} x_{s, \alpha}$.
By \cite[Theorem 4.7]{Kallenberg20} a solution $x \in \Real^{\SA}$ to (S1)-(S3) corresponds one-to-one to an MR policy $\sched$ for $\cM$ such that $x_{s,\alpha} = \freq{\sched}{s,\alpha}$ for all $(s,\alpha) \in \SA$.

\section{Measuring the Quality of a Predictor}
\label{sec:Quality}

The goal of this section is to measure how well reaching a set of states $C$ predicts that a set of terminal states $E$ will be reached in an MDP $\cM$.
For this, we consider well-known measures for binary classifiers from statistical analysis.
To apply such measures, the non-determinism in $\cM$ needs to be resolved.
However, without any assumptions about the resolution of the non-determinism, we propose to consider an average over possible resolutions of the non-determinism.

\subsection{Averaging over Policies}
To be able to compute this average, one has to choose which policies to consider and how to weigh them.
In this paper, we use a uniform measure over MR policies. 
Since we consider reachability properties in particular, the choice of MR policies is justified, as any policy in an EC-free MDP for these properties can be approximated by MR policies \cite{Kallenberg20}.
Furthermore, non-determinism can be used to model uncertainty in the transition probabilities as also represented in interval-Markov chains (see, e.g., \cite{KozineU02,SenVA-IntervalMC06}).
In this case, actions can be used to model the extremal transition probabilities.
Randomizing over such actions then allows for any possible concrete distribution over successor states. 

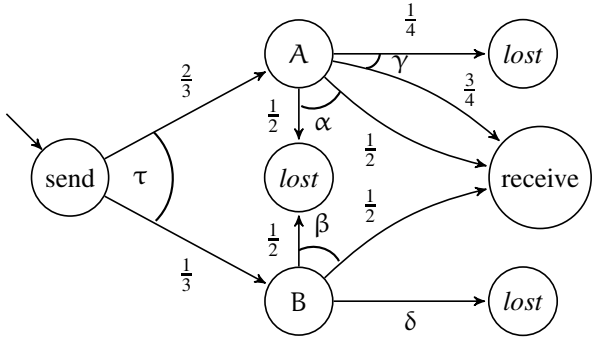
\begin{figure}[t]
	\centering
	\resizebox{0.45\textwidth}{!}{
		\begin{tikzpicture}[->,>=stealth',shorten >=1pt,auto,node distance=0.5cm, semithick]
			
			\node[scale=1, state] (sender) {send};
			\node[scale=1, state] (lost1) [right = 2 of sender]{$\mathit{lost}$};
			\node[scale=1, state] (b) [below = 0.7 of lost1] {$B$};
			\node[scale=1, state] (a) [above = 0.7 of lost1] {$A$};
			
			\node[scale=1, circle, draw] (receiver) [right = 2 of lost1]{receive};
			\node[scale=1, state] (lost2) [right = 2 of a]{$\mathit{lost}$};
			\node[scale=1, state] (lost3) [right = 2 of b]{$\mathit{lost}$};	
			
			\draw[<-] (sender) --++(-.85,0.85);
			
			\draw[color=black, ->] (sender) edge node [anchor=center, pos=0.3] (init1) {} node [pos=0.5, above=0.1] {$\frac{2}{3}$} (a);
			\draw[color=black, ->] (sender) edge node [anchor=center, pos=0.3] (init2) {} node [pos=0.5, below=0.1] {$\frac{1}{3}$} (b);

			\draw[color=black, ->] (a) edge [bend right=15] node [anchor=center, pos=0.1] (a1) {} node [pos=0.3, below=0.1] {$\frac{1}{2}$} (receiver);
			\draw[color=black, ->] (a) edge node [anchor=center, pos=0.3] (a2) {} node [pos=0.5, left=0.1] {$\frac{1}{2}$} (lost1);
			
			\draw[color=black, ->] (a) edge [bend left=15] node [anchor=center, pos=0.15] (aa1) {} node [pos=0.8, above=0.01] {$\frac{3}{4}$} (receiver);
			\draw[color=black, ->] (a) edge node [anchor=center, pos=0.3] (aa2) {} node [pos=0.5, above=0.1] {$\frac{1}{4}$} (lost2);
			
			\draw[color=black, ->] (b) edge [bend left=15] node [anchor=center, pos=0.1] (b1) {} node [pos=0.3, above=0.1] {$\frac{1}{2}$} (receiver);
			\draw[color=black, ->] (b) edge node [anchor=center, pos=0.3] (b2) {} node [pos=0.5, left=0.1] {$\frac{1}{2}$} (lost1);

			\draw[color=black, ->] (b) edge node [pos=0.5,below] {$\delta$} (lost3) ;
			
			\draw[color=black, thick, -] (init1.center) edge [bend left=45] node [pos=0.5, left=0.2] {$\tau$} (init2.center);
			\draw[color=black, thick, -] (a1.center) edge [bend left=45] node [pos=0.5, below=0.01] {$\alpha$} (a2.center);
			\draw[color=black, thick, -] (aa2.center) edge [bend left=45] node [pos=0.5, right=0.1] {$\gamma$} (aa1.center);
			
			\draw[color=black, thick, -] (b2.center) edge [bend left=45] node [pos=0.5, above=0.01] {$\beta$} (b1.center);
		\end{tikzpicture}
	}
	\caption{An MDP $\cM$ depicting a communication network.}
	\label{fig:example-quality}
\end{figure}

\begin{example}
\label{ex:quality}
As an example, consider a communication network in which a message is sent by a sender to a receiver via various network nodes. 
Each node forwards the message to other nodes in a randomized fashion. 
We only know the successors of each network node as well as upper and lower bounds on the respective probability and on the probability that the message is lost between two nodes.
Suppose we have the suspicion that some sets of nodes are faulty and are the reason for many message losses. In order to check this claim,
we want to measure how well reaching such a set of nodes serves as a predictor for a message loss by considering the average over possible resolutions of the non-determinism.
A very simple example of such a network where actions model the known probability bounds is given in Figure \ref{fig:example-quality}.
\Ende
\end{example}

Given an MDP $\cM = (S, \Act,P,\init)$ with terminal states $T\subseteq S$, the set of all
 MR policies $\sched$ of $\cM$ can be described by the set of vectors
 \begin{align*}
	 \mathfrak{P} = \{x \in [0,1]^{\StAct} \mid  \!\!\!\!\sum_{\alpha \in \Act(s)}\!\!\!\! x_{s, \alpha} = 1  \ \ \ \text{ for all $s\in S\setminus T$}\}.
 \end{align*}
	The component  $x_{s, \alpha}$ of $x \in \mathfrak{P}$ expresses the probability that the corresponding policy chooses $\alpha$ in $s$.
	$\mathfrak{P}$ is a product of $\lvert S \!\setminus\! T \rvert$-many regular simplices.
	For $s \in S \!\setminus\! T$ let $n(s) = \lvert \Act(s) \rvert-1$.
	Then, $\mathfrak{P} =  \prod_{s \in S} \Delta^{n(s)}$, where $\Delta^m=\{y\in [0,1]^{m+1} \mid \sum_{i=1}^{m+1} y_i =1\}$ is the regular $m$-simplex of dimension $m$ in $\mathbb{R}^{m+1}$ for given $m \in \Nat$.
	In turn, the dimension of $\mathfrak{P}$ is $\lvert \StAct \rvert - \lvert S\setminus T \rvert$.
	
	\begin{example}
	\label{ex:polytope}
	For the MDP $\cM$ depicted in Fig. \ref{fig:example-quality} considering the state-action pairs $(\init, \tau), (A,\alpha),(A,\gamma),(B,\beta),(B,\delta)$ we get $\mathfrak{P} = \{(1,p,1{-}p, q,1{-}q) \mid p,q\in [0,1]\}$. We can identify $\mathfrak{P}$ with the product of simplices $\{1\}\times[0,1]\times[0,1]$ or simply $[0,1]^2$.
	\Ende
	\end{example}

	Using the Lebesgue measure, we get a way to uniformly average over all MR policies.
	As mentioned above, we want to apply this to some quality measure for binary classifiers.
	For a quality measure $Q^x(C)$ for the predictor $C \subseteq S$ depending on a policy $x\in \mathfrak{P}$, the average of the quality measure is
	\[
		{\int_{\mathfrak{P}} Q^x(C) \,\mathrm d x} \,\, / \,\,{\int_{\mathfrak{P}} 1\, \mathrm d x}.
	\]	
The computation of $V(\mathfrak{P})=\int_{\mathfrak{P}} 1\, \mathrm d x$ can be done in polynomial time since $\mathfrak{P}$ is the product of standard simplices as described above.
Then, by \cite{Simplex-Volume} we have
\begin{align*}
	\int_{\mathfrak{P}}\!\! 1 \, \mathrm dx & = V(\prod_{s \in S} \Delta^{n(s)}) = \prod_{s \in S} V(\Delta^{n(s)}) = \prod_{s \in S} \frac{1}{n(s)!},
\end{align*}
which is computable in polynomial time.
In fact, integrating any polynomial of fixed degree over $\mathfrak{P}$ can also be done in polynomial time \cite{BBdLKV-integrating-over-simplex-2011}.

\setlength{\tabcolsep}{1mm}
\begin{figure}[t]
	\centering
	\begin{tabular}{c||c|c}
		& $\lozenge\Eff$ & $\neg \lozenge\Eff$ \\
		\hline
		\hline
		$\lozenge\Cause$& True positive & False positive  \\
		& $\tp^\sched_{\cM} \! =\Pr^\sched\!(\lozenge \Cause \wedge \lozenge \Eff)$ & $\fp^\sched_{\cM} \! =\Pr^\sched\!(\lozenge \Cause \wedge \neg \lozenge \Eff)$ \\
		\hline
		$\neg \lozenge \Cause$ & False negative   & True negative   \\
		&$\fn^\sched_{\cM} \! =\Pr^\sched\!(\neg \lozenge \Cause \wedge \lozenge \Eff)$ & $\tn^\sched_{\cM} \! =\Pr^\sched\!(\neg \lozenge \Cause \wedge \neg \lozenge \Eff)$ \\
	\end{tabular}
\caption{The confusion matrix for the prediction of $\Eff \subseteq S$ by $\Cause \subseteq S$ for a given policy $\sched$.}
	\label{fig:confusion}
\end{figure}

\subsection{Quality Measures}
Given a single policy $\sched$ for $\cM$, we consider the so-called \emph{confusion matrix} \cite{Powers-fscore} for binary classifiers as depicted in Figure \ref{fig:confusion}.
In statistical analysis, various measures for the quality of a classifier using the entries of the confusion matrix have been studied. 
Important measures are
\begin{align*}
	&\precision^\sched(\Cause) = \frac{\tp^\sched}{\tp^\sched + \fp^\sched}, \
	\recall^\sched(\Cause) = \frac{\tp^\sched}{\tp^\sched + \fn^\sched}, \\
	&\fscore^\sched\!(\Cause) \! = \! \frac{2}{\frac{1}{\precision^\sched\!(\Cause)} +\frac{1}{\recall^\sched\!(\Cause)} }\! =\! \frac{2\tp^\sched} {2\tp^\sched\!+\fp^\sched\!+\fn^\sched}.
	\end{align*}
Intuitively, the $\precision^\sched(\Cause)$ measures the probability that the prediction is indeed true after $\Cause$ is reached.
The $\recall^\sched(\Cause)$, on the other hand, expresses the probability that reaching $\Eff$ was preceded by reaching $\Cause$ under the condition that $\Eff$ has indeed been reached.
The $\fscore^\sched(\Cause)$ is the harmonic mean of precision and recall. A high f-score hence indicates that after reaching the predictor $\Cause$, the probability to reach $\Eff$ is relatively high and at the same time relatively many executions leading to $\Eff$ pass through $\Cause$ before.
An example for a more complex quality measures -- similar in spirit to the f-score \cite{chicco2020advantages} -- is Matthews correlation coefficient given by $\mcc^\sched(\Cause) = $
\begin{align*}
	&\frac{\tp^\sched \cdot \tn^\sched - \fp^\sched \cdot \fn^\sched}{\sqrt{(\tp^\sched + \fp^\sched) \cdot(\tp^\sched + \fn^\sched) \cdot (\tn^\sched + \fp^\sched) \cdot (\tn^\sched + \fn^\sched)}}.
\end{align*}

\paragraph*{Computing average quality measures.}
In order to compute the average quality measures over the polytope $\mathfrak{P}$ of MR policies, we have to express the measures in terms of the policies.
As the considered quality measures depend on the confusion matrix, the task hence is to express the entries of the confusion matrix in terms of the vectors $x\in \mathfrak{P}$

To express these values, we take a small detour via a model transformation such that we can distinguish whether $\Cause$ has occurred or not.
We define the \emph{two-copy MDP} $\twoMDP{\cM}{\Cause}$ of $\cM$ with respect to $\Cause$ in the following way:
\begin{itemize}
	\item $\twoMDP{\cM}{\Cause}$ consists of two copies of $\cM$, namely $\cM_0$ and $\cM_1$,
	\item the initial state is $\init_0$ of the first copy,
	\item whenever a state $c_0 \in \Cause_0$ is reached in the first copy there is exactly one action which transitions to the corresponding state $c_1$ in the second copy with probability $1$.
\end{itemize}
A policy $\sched$ of the original MDP $\cM$ can be interpreted as a policy $\usched$ of the two-copy MDP $\twoMDP{\cM}{\Cause}$ by  mimicking the behavior in both copies ignoring the switch to the second copy. We denote the copies of the states in $\Cause$ and $\Eff$ in $\cM_0$ and $\cM_1$ by $\Cause_0$, $\Cause_1$, $\Eff_0$, and $\Eff_1$, respectively.
Now, we can  describe the entries of the confusion matrix for $\cM$ directly by reachability probabilities in $\twoMDP{\cM}{\Cause}$.
\begin{lemma}
\label{lem:confusion}
	For an MR policy $\sched$ of $\cM$ also viewed as a policy for $\twoMDP{\cM}{\Cause}$ the following holds
	\begin{align*}
		\tp^\sched_\cM &= \Pr^\sched_{\twoMDP{\cM}{\Cause}}(\lozenge \Eff_1), \quad \fn^\sched_\cM = \Pr^\sched_{\twoMDP{\cM}{\Cause}}(\lozenge \Eff_0),  \\
		\fp^\sched_\cM &=  \!\!\sum_{c \in \Cause_0}    \!\!  \Pr^\sched_{\twoMDP{\cM}{\Cause}}(\lozenge c) \cdot (1 - \Pr^\sched_{\twoMDP{\cM}{\Cause}}(\lozenge \Eff_1)), \\
		\tn^\sched_\cM & = 1 - \tp^\sched_\cM - \fp^\sched_\cM - \fn^\sched_\cM.
	\end{align*}
\end{lemma}
\begin{proof}
	Only the equation for $\fp^\sched_\cM$ must be proven since the other equations follow by construction.
	 The probability that $c\in \Cause_0 $ is visited and afterwards $\neg \lozenge \Eff_1$
	holds is $\Pr^{\sched}_{\twoMDP{\cM}{\Cause}} (\lozenge c) \cdot (1 - \Pr^\sched_{\twoMDP{\cM}{\Cause}, c}(\lozenge \Eff_1))$. As the set $\Cause_0$ can only be reached once  in $\twoMDP{\cM}{\Cause}$, 
	the equation for $\fp^\sched_{\cM}$ follows by adding the probabilities of these disjoint events.
	\end{proof}
	We denote the state space of $\twoMDP{\cM}{\Cause}$ by $S'$ and the set of terminal states by $T'$.
Now, we describe reachability probabilities between states in $\twoMDP{\cM}{\Cause}$ in terms of distributions given by MR policies.
For  $s, t \in S'$ and an MR policy $\sched$ of $\cM$ let $a^\sched_{s,t} = \Pr^\sched_{s}(\lozenge t)$ be the probability to eventually reach $t$ from $s$ under $\sched$.
So, $a^\sched_{s,s} = 1$ for all $s \in S'$.
Using the representation 
of MR policies as vectors $x\in \mathfrak{P}$, we then have for non-terminal states $s\in S'\setminus T'$ and $t\not=s$,
\begin{align}
	\label{reach}\tag{reach}
	a^x_{s,t} = \!\!\!\sum_{\alpha \in \Act(s)}\!\!\! x_{s, \alpha} \cdot \sum_{u \in S} P(s, \alpha, u) \cdot a^x_{u,t}.
\end{align}
In fact, after determining the states $s'$ from which $t$ is not reachable in the Markov chain induced by $x$ and  $\twoMDP{\cM}{\Cause}$ and setting the corresponding variables $a^x_{s',t}$ to $0$, the equation system has a unique solution (see \cite{BaierK2008}).
Given a rational-valued MR policy $x\in \mathfrak{P}$ the reachability probabilities can hence be derived in polynomial time by solving the linear equation system.

However, we want to compute the average of quality measures by taking an integral over the polytope $\mathfrak{P}$ of all MR policies. As we use the standard Lebesgue measure, the boundary 
of this polytope has measure $0$. So, for our purpose, it is sufficient to express the quality measures as a function of the policy $x\in \mathfrak{P}$ on the interior of $\mathfrak{P}$.

\begin{proposition}
The values $a^x_{s,t} $ for $s,t\in S'$ are rational functions in $x$ on the interior of $\mathfrak{P}$. The degree of denominator and enumerator of these rational functions is $\leq 2 |S|$.
\end{proposition}

\begin{proof}
For a fixed $t \in S'$, the set $A_0$ of states $s'$ from which $t$ is not reachable in the Markov chain induced by $x$ and $\twoMDP{\cM}{\Cause}$ depends on which entries of $x$ are $0$. 
In the interior of $\mathfrak{P}$,  no vector has a $0$-entry. So, $A_0$ is independent of $x$ and the corresponding variables can be set $0$.
As the resulting equation system with at most $|S'|$-many variables has a unique solution (see \cite{BaierK2008}),
this solution can be expressed using fractions of determinants by Cramer's rule. The determinants are polynomials in the coefficients of the linear equations of degree at most 
$|S'|=2|S|$ and the variables $x_{s,\alpha}$ appear only linearly in these coefficients. 
\end{proof}
The polynomials in the resulting rational function contain exponentially many monomials in general. 
Nevertheless, we know that the values $a^x_{s,t} $  are rational functions in $x$ on the interior of $\mathfrak{P}$ that only take values between $0$ and $1$.

\paragraph{Practical considerations for averages.}

If the transition relation is sparse, which is often the case for models with large state space, results on the efficient computation of symbolic determinants apply \cite{DBLP:journals/cc/KaltofenV05,DBLP:conf/meco/DuriqiSSL24}.
Then, an exact  representation of the rational functions $a^x_{s,t} $ can be obtained efficiently.

By Lemma \ref{lem:confusion}, all entries of the confusion matrix under policy $x\in\mathfrak{P}$
can be expressed in terms of $a^x_{s,t}$ by simple arithmetic. Furthermore, the prominent measures for the quality of a predictor such as 
precision, recall, and f-score are linear rational functions in these entries. But also more complex measures such as the Matthews correlation coefficient 
are still relatively simple functions in the entries of the confusion matrix. So, all of these quality measures can be expressed 
as functions $Q(a^x)$ where $a^x$ is the vector containing the values $a^x_{s,t}$. 
Unfortunately, an exact evaluation of the integral
$\int_{\mathfrak{P}} Q(a^x) \,\mathrm dx$ is nevertheless typically out of reach.
But the quality measures take values in $[0,1]$ or $[-1,1]$ in the case of the MCC. Together with the fact, that the resulting function $Q(a^x)$ are smooth functions
on the interior of $\mathfrak{P}$, standard approaches such as Monte Carlo integration can be used to approximate the integral \cite{NumRecipies2007}. Note, that we do not need an explicit representation 
of the functions $a^x_{s,t}$ when we sample policies $x\in \mathfrak{P}$ as the values $a^x_{s,t}$ can then be computed in polynomial time.

\begin{example}
\label{ex:computation}
Let us apply average quality measures to the network MDP $\cM$ depicted in Figure \ref{fig:example-quality}.
First, we consider $A$ as the predictor. Here, we identify policies with pairs $(p,q)\in[0,1]^2$ as in Example \ref{ex:polytope} where $p$ is the probability to choose $\alpha$ in $A$
and $q$ the probability to choose $\beta$ in $B$. For a policy $x=(p,q)$, we get
\begin{align*}
	\tp^x_\cM &= \frac16(1+ p), &\fp^x_\cM &= \frac16(3- p), \\
	\tn^x_\cM &= \frac16 q, &\fn^x_\cM &= \frac16(2- q).
\end{align*}
As the volume of $\mathfrak{P}$ is $1$ in this case, we obtain the average f-score, for example, as
\begin{align*}
&\int_{[0,1]^2} \fscore^{x}(\{A\}) \,\mathrm dx = \int_{[0,1]^2}  \frac{2\tp^\sched} {2\tp^\sched+\fp^\sched+\fn^\sched} \,\mathrm dx\\
=& \int_0^1 \int_0^1 \frac{2+2p}{7+p -q} \, \mathrm dp \, \mathrm dq \approx 0.43 .
\end{align*}
Analogously, we obtain
\[
\int_{[0,1]^2} \fscore^{x}(\{B\}) \,\mathrm dx = \int_0^1 \int_0^1 \frac{4-2q}{5+p -q} \, \mathrm dp \, \mathrm dq  \approx 0.60.
\]
So, according to the f-score, reaching $B$ is on average a better predictor for a message loss than reaching $A$.
\Ende
\end{example}

\begin{remark}
	If a quality measure can be written as a linear rational function in terms of the entries of the confusion matrix, then its minimal or maximal value can be computed with the techniques presented in \cite{LMCS24}.
	This is e.g. the case for precision, recall and f-score.
	For this a standard model transformation is performed which collapses end components \cite{deAlfaro1997, Alfaro-CONCUR99} while preserving relevant reachability probabilities (c.f. \cite[III.B]{QEST08} for a compact description).
	Afterwards, possible combinations of reachability probabilities can be expressed in terms of a linear constraint system for state-action pair frequencies. For linear rational quality measures, the resulting optimization problem can be solved in polynomial time.
	For more complex measures like Matthews correlation coefficient, more general optimization problems arise.
\end{remark}

\section{Probability-Raising Policies and the Causal Volume of a Predictor}
\label{sec:PR-policy}

In this section we investigate, whether in an MDP $\cM$ a given predictor $\Cause$ has a probabilistic cause-effect relation with the undesired event $\Eff$.
The idea that a cause for an event serves as a good predictor comes very natural and the usage of probabilistic causes for predictions has already been considered in \cite{ISSE22} for Markov chains.
For MDPs we take inspiration from the notion of probability-raising causes from \cite{LMCS24} to define two variants of \emph{probability-raising (PR) policies}, which are, in simple terms, witnessing a probability-raising condition.
Since the focus of this paper is to investigate the quality of a predictor, we introduce a measure which considers the relative amount of possible MR policies that witness a PR condition.
Afterwards we study the complexity of deciding the existence of such PR policy for a given predictor.
The proofs in this section will only be sketched here, but can be found in the appendix \ref{sec:ap-PR-policy}.

\begin{definition}[Probability-raising policy]
	\label{def:PR-policy}
	Given $\cM$, $\Effect$ and $\Cause$, a policy $\sched$ of $\cM$ is a \emph{global probability-raising policy} (GPR policy) for $\Cause$ and $\Eff$ in $\cM$ if the following conditions \condR and \condG hold
	\begin{description}
		\item[(R)] \hspace{26pt}$\Pr^\sched(\lozenge \Cause) > 0$,
		\item[(G)] \hspace{26pt}$\Pr^\sched(\lozenge \Eff \mid \lozenge \Cause) > \Pr^\sched(\lozenge \Eff)$. \hfill (GPR)
	\end{description}
	$\sched$ is a \emph{strict probability-raising policy} (SPR policy) for $\Cause$ and $\Eff$ in $\cM$ if \condR and the following condition \condS hold
	\begin{description}
		\item[(S)] For all $c \in \Cause$ with $\Pr^\sched((\neg \Cause) \until c) > 0$:
		\begin{align}
			\label{SPR}
			\Pr^\sched(\lozenge \Eff \mid (\neg \Cause) \until c) > \Pr^\sched(\lozenge \Eff).
			\tag{SPR}
		\end{align}
	\end{description}
\end{definition}

There is an implication from strict to global, since the conditional probability in (GPR) is a weighted sum over the conditional probabilities in (SPR) (also cf. \cite{LMCS24}).
Similarly, for singletons $\{c\}$ the equivalence of $\lozenge c$ and $(\neg c) \until c$ means that in such cases the strict and global PR conditions coincide.

Under a PR policy $\sched$ there is a causal relationship between the predictor $\Cause$ and the predicted event $\Eff$.
Namely, whenever $\Cause$ is reached under a PR policy $\sched$, the probability of reaching $\Eff$ is raised.

\begin{example}
	\label{ex:general-EPR}
	For an example we consider the network MDP $\cM$ from Figure \ref{fig:example-quality} as before.
	Here we consider $\Cause = \{B\}$ as a predictor for $\Eff = \{\mathit{lost}\}$.
	For the MD policy $\sched$ choosing $\gamma$ in $A$ and $\beta$ in $B$ we have
	\begin{align*}
		\Pr^\sched(\lozenge \mathit{lost} \mid \lozenge B) = \frac{1}{2} > \frac{1}{3} \cdot \frac{1}{2} + \frac{2}{3} \cdot \frac{1}{4} = \frac{1}{3} = \Pr^\sched(\lozenge \mathit{lost})
	\end{align*}
	and thus $\sched$ satisfies (GPR) (and (SPR)) and  constitutes a PR policy.
	However, there are  policies which do not satisfy (GPR), e.g., the MD policy $\usched$ choosing $\alpha$ in $A$ and $\beta$ in $B$.~\Ende
\end{example}

\subsection{Causal volumes}
\label{sec:causal-volume}

With the definition of probability-raising policy in mind the question now arises, in how many cases such a causal relation holds between the predictor and the undesired outcome.
For this we consider the relative portion of PR policies among all possible MR policies.
Recall the polytope $\mathfrak{P}$ of distributions of MR policies in $\cM$ from Section \ref{sec:Quality}.
Using the reachability matrix corresponding to a given policy $x \in \mathfrak{P}$ from \eqref{reach} we express the conditions \condR, \condS and \condG from Def. \ref{def:PR-policy} by
\begin{align}
	\sum_{c \in \Cause} a^x_{\init, c} &> 0,
	\tag{r}\label{r} \\
	\sum_{\eff \in \Eff}\!\!\! a^x_{c, \eff}  > \sum_{\eff \in \Eff}\!\!\! a^x_{\init, \eff} \ \ \ &\text{for all} \ c \in \Cause \ \text{with} \ a^x_{\init, c} > 0,
	\tag{s}\label{s}\\
	\tp^x \cdot \tn^x &- \fp^x \cdot \fn^x > 0.
	\label{g}\tag{g}
\end{align}

We define the sets of MR policies which are SPR (or respectively GPR) policies by
\begin{center}
	$\mathfrak{P}_{SPR} = \{x \in \mathfrak{P} \mid x \ \text{satisfies \eqref{reach}, \eqref{r} and \eqref{s}} \},$ \\
	$\mathfrak{P}_{GPR} = \{x \in \mathfrak{P} \mid x \ \text{satisfies \eqref{reach}, \eqref{r} and \eqref{g}} \}.$
\end{center}

The three considered sets have the same dimension:

\begin{theorem}
	\label{thm:full-dimension}
	For an MDP $\cM$, a set of terminal states $\Eff$ and a predictor set $\Cause$ for which a memoryless SPR (resp. GPR) policy exists, the sets $\mathfrak{P}$ and $\mathfrak{P}_{SPR}$ (resp. $\mathfrak{P}_{GPR}$) have the same dimension.
\end{theorem}
\begin{proof}[Proof sketch]
	The claim follows from the fact that for any $x \in \mathfrak{P}_{SPR}$ (resp. $\mathfrak{P}_{GPR}$) there is an $\varepsilon > 0$ such that the $\varepsilon$-neighborhood of $x$ is completely contained in $\mathfrak{P}$.
\end{proof}

Recall the set of all terminal states $T \subset S$ of $\cM$. With this, we can now define the following \emph{Volumes} with \mbox{$\lvert \StAct \rvert - \lvert S \setminus T \rvert$}-dimensional Lebesgue integrals:
\begin{align*}
	V(\mathfrak{P}) &\eqdef \int_{\mathfrak{P}}1 \ d x, \\
	V(\mathfrak{P}_{SPR}) \eqdef \int_{\mathfrak{P}_{SPR}}\!\!\!\! 1 \ d &x,  \quad V(\mathfrak{P}_{GPR}) \eqdef \int_{\mathfrak{P}_{GPR}}\!\!\!\! 1 \ d x.
\end{align*}

\begin{definition}
	Let $\cM$ be an MDP, $\Eff \subset S$ a set of terminal states and $\Cause \subseteq S \setminus \Eff$.
	We define the \emph{strict causal volume} and resp. \emph{global causal volume} of $\Cause$ for $\Eff$ as
	\begin{align*}
		\operatorname{sV}(\Cause) = \frac{V(\mathfrak{P}_{SPR})}{V(\mathfrak{P})} \quad \text{and} \quad
		\operatorname{gV}(\Cause) = \frac{V(\mathfrak{P}_{GPR})}{V(\mathfrak{P})}.
	\end{align*}
\end{definition}

These causal volumes now express the fraction of MR policies which constitute SPR or GPR policies.
Having an estimate of the causal volume of a predictor is an additional information about its quality.
Consider, for example, a large network in which a message is sent from one node to another.
The information that a component is a probabilistic cause for message to be lost in a lot of cases gives a good reason to predict the message loss whenever the component is part of the communication.

\begin{example}
	For a smaller network example we again consider the MDP from Figure \ref{fig:example-quality} where the undesired outcome $\Eff = \{\mathit{lost}\}$ is to be predicted by $\Cause = \{B\}$.
	Here, we conclude from Example \ref{ex:general-EPR} that PR policies exist.
	Since $\Cause$ is a singleton the sets for SPR and GPR policies coincide and we have
	\begin{align*}
		\mathfrak{P}_{GPR} = \{x \in [0,1]^\StAct \mid x_{A, \alpha} < 1 \ \text{or} \ x_{B, \beta} < 1\},
	\end{align*}
	since only the MD policy choosing $\alpha$ and $\beta$ does not constitute a PR policy.
	Thus, we also have
		$\operatorname{sV}(\Cause) = \operatorname{gV}(\Cause) = 1$
	since almost all MR policies are both SPR and GPR policies.
	This means that from the perspective of probabilistic causality, the event $\lozenge B$ is a good predictor for $\lozenge \mathit{lost}$.
	\Ende
\end{example}

As mentioned in Section \ref{sec:Quality} the computation of $V(\mathfrak{P})$ can be done in polynomial time by representing $\mathfrak{P}$ as a product of simplices.
However, in general the exact computation of the volume of a fully-dimensional polytope given in halfspace-representation such as $V(\mathfrak{P}_{SPR})$ is $\#\PTIME$-hard \cite{BW1991}.
Moreover, it has been shown that for any polynomial-time approximation there is a minimal gap between upper and lower bounds depending on the dimension of the problem \cite{Barany1987}.
With respect to these restrictions there are still efficient tools for the exact computation of polynomials over polytopes \cite{deLoera2013}.
For $\mathfrak{P}_{GPR}$ exact integration is even harder, since it is not a polytope by equation \eqref{g}.
However, since \eqref{g} is the only non-linear function restricting $\mathfrak{P}_{GPR}$ this favors the usage of Monte Carlo integration algorithms \cite{NumRecipies2007} to approximate $V(\mathfrak{P}_{GPR})$.

\subsection{Checking Probability-Raising Policies}
\label{sec:PR-check}

While computing the causal volume for a given predictor and outcome is a difficult problem, we now want to address the existential query for probability-raising policies:

\begin{center}
	Given an MDP $\cM$, a set of terminal states $\Eff \subseteq S$ and a predictor set $\Cause \subseteq S \setminus \Eff$, is there a strict (resp. global) probability-raising policy for $\Cause$ and $\Eff$ in $\cM$?
\end{center}

For this existence check we use a model transformation similar to the two-state MDP of $\cM$ (Sec. \ref{sec:Quality}).
This results in an end-component free MDP in which both $\Cause$ and $\Eff$ can only be visited once and there are exactly \emph{four} terminal states corresponding to the entries of the confusion matrix.
So, we can express these entries as reachability probabilities of single terminal states.
We use the abbreviations from Figure \ref{fig:confusion}.

\begin{definition}[Canonical MDP]
	\label{def:canonical}
	We transform the original MDP $\cM$ to the \emph{canonical MDP} $\canMDP{\cM}{\Cause}$ in the following way:
	\begin{itemize}
		\item[(i)] All outgoing transitions from states $c \in \Cause$ are deleted and instead two fresh actions $\alpha_{\min}$ and $\alpha_{\max}$ are added.
		They transition to a new state $\TP$ with the minimal and resp. maximal probability to reach $\Eff$ from $c$.
		With the remaining probability they transition to a new state $\FP$.
		\item[(ii)] Collapse the maximal end-components (MECs) $\cE$ of the resulting MDP into single states $s_{\cE}$ by taking the MEC-quotient, see e.g. \cite{Alfaro-CONCUR99}.
		\item[(iii)] Collapse the states of $\Eff$ to a fresh state $\FN$ and other terminal states to the fresh state $\TN$.
		States $s_{\cE}$ representing MECs get the additional transition $P(s_{\cE} , \tau, \TN) = 1$.
		The resulting MDP is $\canMDP{\cM}{\Cause}$.
	\end{itemize}
	In $\canMDP{\cM}{\Cause}$ we consider $\Eff' = \{\TP, \FN\}$.
	\Ende
\end{definition}

Intuitively, the first step (i) of the transformation ensures that each state $c \in \Cause$ is visited at most once, while preserving all possible values for positive predictions.
In the second (ii) and third step (iii) the transformation gets rid of end-components by collapsing them into single states.
Instead, the fresh action $\tau$ corresponds to the case that in the original MDP $\cM$ a policy $\sched$ realizes a true negative by staying in an end-component $\cE$ indefinitely.
The soundness of the canonical MDP for a given set $\Cause$ with respect to \condG and \condS follows from the results of \cite{LMCS24}.

\begin{lemma}[$\canMDP{\cM}{\Cause}$ preserves Confusion Matrix]
	\label{lem:preserving confusion matrix}
	Given MDP $\cM$, set of terminal states $\Eff \subset S$ and $\Cause \subseteq S \!\setminus\! \Eff$, for each policy $\sched$ of $\cM$ there is a policy $\usched$ of $\canMDP{\cM}{\Cause}$ and vice versa such that
	\begin{center}
		$\tp^\sched_\cM = \Pr^\usched_{\canMDP{\cM}{\Cause}}(\lozenge \TP), \qquad \fp^\sched_\cM = \Pr^\usched_{\canMDP{\cM}{\Cause}}(\lozenge \FP),$ \\
		$\fn^\sched_\cM = \Pr^\usched_{\canMDP{\cM}{\Cause}}(\lozenge \FN), \qquad \tn^\sched_\cM = \Pr^\usched_{\canMDP{\cM}{\Cause}}(\lozenge \TN).$
	\end{center}
\end{lemma}

\begin{corollary}[$\canMDP{\cM}{\Cause}$ preserves Probability-Raising]
	\label{cor:preserving PR}
	Given an MDP $\cM$, a set of terminal states $\Eff \subset S$ and a set $\Cause \subseteq S \setminus \Eff$, a policy $\sched$ of $\cM$ satisfies \condS (resp. \condG ) for $\Cause$ and $\Eff$ in $\cM$ iff the corresponding policy $\usched$ from Lemma \ref{lem:preserving confusion matrix} satisfies \condS (resp. \condG ) for $\Cause$ and $\{\TP, \FN\}$ in $\canMDP{\cM}{\Cause}$.
\end{corollary}

\subsubsection{Checking for SPR Policies}
\label{sec:check-SPR-policy}

We now consider the following problem: Given an MDP $\cM$ with a set of terminal states $\Eff \subset S$ and a predictor set $\Cause \subseteq S \setminus \Eff$, is there a policy $\sched$ of $\cM $ such that (SPR) holds for $\sched$?
By the soundness of the canonical MDP w.r.t. the PR conditions (Cor. \ref{cor:preserving PR}) and the confusion matrix (Cor. \ref{cor:preserving PR}) we assume $\cM = \canMDP{\cM}{\Cause}$.

For an SPR policy there needs to be a balancing between states $c \in \Cause$ with a high value $p_{c, \max}$ and states $c' \in \Cause$ with a low value $p_{c', \max}$.
We will provide a characterization of SPR policies using this idea.
For this, we consider the maximal probability to reach $\TP$ among all cause states: 
\begin{center}
	$p^\star = \max_{c \in \Cause} p_{c, \max}$
\end{center}
From $\cM$ to $\cM^\star$ we only change the actions in states $c \in \Cause$.
For each $c$ the only enabled action in $\cM^\star$ is $\delta$ with
\begin{center}
	 \mbox{$\ \ \ \: P'(c, \delta, \TP) = p^\star$} and \mbox{$P'(c, \delta, \FP) = 1 - p^\star$}
\end{center}
So, $\cM^\star$ behaves as $\cM$, but when a state $c \in \Cause$ is reached the single enabled action leads to $\TP$ with $p^\star$.
By construction, any policy $\usched$ of $\cM^\star$ corresponds to a policy $\sched$ of $\cM$.

\begin{lemma}[Characterization of SPR Policies]
	\label{lem:char-SPR-policy}
	For an MDP $\cM$ with set of terminal states $\Eff \subset S$ and set of states $\Cause \subseteq S \setminus \Eff$ there is an SPR policy $\sched$ for $\Cause$ and $\Eff$ in $\cM$ iff $\Pr^{\min}_{\cM^\star}(\lozenge \Eff) < p^\star$.
\end{lemma}

\begin{theorem}[Complexity for Checking SPR policy]
	\label{thm:SPR policy in polytime}
	Given an MDP $\cM$ with set of terminal states $\Eff \subset S$ and set of states $\Cause \subseteq S \setminus \Eff$ the existence of an SPR policy for $\Cause$ and $\Eff$ in $\cM$ can be decided in $\PTIME$.
	Moreover, a corresponding finite-memory (randomized) policy $\sched$ can be computed in polynomial time.
\end{theorem}
\begin{proof}
	We first note that we can transform $\cM$ to $\canMDP{\cM}{\Cause}$ in polynomial time.
	We can now rely on the characterization of Lemma \ref{lem:char-SPR-policy}.
	So, by further transforming $\canMDP{\cM}{\Cause}$ to $\canMDP{\cM}{\Cause}^\star$ we can check whether $\Pr^{\min}_{\canMDP{\cM}{\Cause}^\star} < p^\star$ in polynomial time \cite{BaierK2008}.
	
	If this check is positive, an MD policy $\tsched$ of $\canMDP{\cM}{\Cause}^\star$ with $\Pr^{\tsched}_{\canMDP{\cM}{\Cause}^\star} = \Pr^{\min}_{\canMDP{\cM}{\Cause}^\star}$ can be derived.
	By \cite{LMCS24}[Theorem 4.19] we can further derive the corresponding finite-memory (randomized) policy $\sched$ for $\cM$ in polynomial time.
\end{proof}

\begin{remark}[Complexity for Singletons]
	\label{rem:complexity singleton EPR}
	The case where $\Cause = \{c\}$ is a singleton can be decided more efficient.
	Note that there does not need to be a balancing between different states of $\Cause$.
	Therefore, it is sufficient to only enable action $\alpha_{\max}$ in $c$.
	Let us call the resulting MDP $\cN$.
	We then only have to check, whether $p_{\max, c} > \Pr^{\min}_{\cN}(\lozenge \Eff)$ as this corresponds to the best case SPR policy.
	\Ende
\end{remark}

\subsubsection{Checking for GPR Policies}
\label{sec:check-GPR-policy}

The next decision problem we consider is: Given an MDP $\cM$, set of terminal states $\Eff$ and predictor set $\Cause$ is there a GPR policy $\sched$ for $\Cause$ and $\Eff$ in $\cM$?
By Lemma \ref{lem:preserving confusion matrix} we assume $\cM = \canMDP{\cM}{\Cause}$ and use this to encode the inequality (GPR) in terms of state-action frequency variables.
For this, we reformulate (GPR) in $\canMDP{\cM}{\Cause}$, using straight forward calculations, to
\begin{center}
	$\Pr^\sched(\lozenge \TP) \cdot \Pr^\sched(\lozenge \TN) - \Pr^\sched(\lozenge \FP) \cdot \Pr^\sched(\lozenge \FN) > 0.$
\end{center}

So, we consider this inequality in terms of frequencies together with the equations (S1)-(S3) (cf. Section \ref{sec:prelim}): 
\begin{align}
	\label{eq:freq-GPR}
	x_\TP \cdot x_\TN - x_\FP \cdot x_\FN > 0.
	\tag{freq-GPR}
\end{align}
\begin{lemma}[Quadratic Program for GPR policy]
	There is a GPR policy for $\Cause$ and $\Eff$ in $\cM$ iff the system of inequalities given by (S1)-(S3) and (freq-GPR) has a solution.
\end{lemma}
\begin{proof}
	Let $\sched$ be a GPR policy for $\Cause$ and $\Eff$ in $\cM$.
	By construction the corresponding frequencies of state-action pairs of $\sched$ are a solution to (S1)-(S3) and (freq-GPR).
	
	Now, assume $x \in \Real^\StAct$ is a solution to (S1)-(S3) and (freq-GPR) and let $\sched$ be an MR policy corresponding to $x$.
	The inequality (freq-GPR) is equivalent to (GPR) in $\canMDP{\cM}{\Cause}$.
	As (freq-GPR) and (S1) hold, we have $x_\TP \cdot x_\TN > 0$ and thus $x_\TP > 0$.
	Since $x_\Cause = x_\TP + x_\FP$ we also get $\Pr^\sched(\lozenge \Cause) > 0$.
	So, $\sched$ is a GPR policy for $\Cause$ and $\Eff$ in $\cM$.
\end{proof}

However, since \eqref{eq:freq-GPR} is a strict inequality we can not directly use quadratic programming to decide the solvability \cite{Vavasis1990}.
We can still give an $\NP$ upper bound for deciding the existence of a GPR policy.

\begin{theorem}[Complexity GPR Policies Check]
	\label{thm:complexity EGPR}
	Deciding whether there is a GPR policy for $\Cause$ and $\Eff$ in $\cM$ can be done in $\NP$.
\end{theorem}
\begin{proof}[Proof sketch]
	The proof uses distinct cases of \eqref{eq:freq-GPR} within the set of possible state-action frequencies.
	It relies on the intermediate value theorem to show that, if there is a possible distribution of frequencies that forces \eqref{eq:freq-GPR} to an equality, then there is an $\varepsilon$-neighborhood in which this becomes an inequality again.
\end{proof}

By the result of \cite{LMCS24}[Theorem 4.19] this existence check results in a finite-memory randomized GPR policy $\sched$ with exactly two memory cells.

\section{Conclusion}

In this paper, we proposed different measures to quantify the quality of predictors in adaptive systems. 
In the stochastic operational model of MDPs we considered predictors (and outcomes) as sets of states of the MDP.
To this end, we proposed two approaches for measuring the effectiveness of the predictor.
First, by averaging over the non-deterministic choices of the MDP we capture the effectiveness of the predictor to foresee undesired events in general.
Second, we measure the fraction of policies that witness probability-raising causality of the predictor to the outcome, where we borrow ideas from recent work on probability-raising causality \cite{LMCS24}.
For the proposed measures we provided insights on the complexity and discussed existing numerical methods for computing the respective measures.

\section*{Acknowledgments}
The authors are supported by the DFG through
	the DFG grant 389792660 as part of TRR~248
	and the Cluster of Excellence EXC 2050/1 (CeTI, project ID 390696704, as part of Germany's Excellence Strategy)
	and by BMBF (Federal Ministry of Education and Research) in DAAD project 57616814 (SECAI, School of Embedded and Composite AI) as part of the program Konrad Zuse Schools of Excellence in Artificial Intelligence.

\newtheorem{innercustomthm}{Theorem}
\newenvironment{ctheorem}[1]
{\renewcommand\theinnercustomthm{#1}\innercustomthm}
{\endinnercustomthm}

\newtheorem{innercustomlem}{Lemma}
\newenvironment{clemma}[1]
{\renewcommand\theinnercustomlem{#1}\innercustomlem}
{\endinnercustomlem}

\newtheorem{innercustomcor}{Corollary}
\newenvironment{ccorollary}[1]
{\renewcommand\theinnercustomcor{#1}\innercustomcor}
{\endinnercustomcor}

\theoremstyle{normal}
\newtheorem{innercustomdef}{Definition}
\newenvironment{cdefinition}[1]
{\renewcommand\theinnercustomdef{#1}\innercustomdef}
{\endinnercustomdef}
	
\appendix
	\section{Appendix}
	\label{sec:ap-PR-policy}
	
	In this section, the detailed proofs of the claims of Section \ref{sec:PR-policy} can be found.
	
	\subsection{Proofs of Section 4.1}
	\label{sec:ap-check-SPR-policy}
	
	Recall the definitions of the sets of MR scheduler
	\begin{align*}
		\mathfrak{P} &= \{x \in [0,1]^{\StAct} \mid  \!\!\!\!\sum_{\alpha \in \Act(s)}\!\!\!\! x_{s, \alpha} = 1  \ \ \ \text{ for all $s\in S\setminus T$}\},\\
		\mathfrak{P}_{SPR} &= \{x \in \mathfrak{P} \mid x \ \text{satisfies \eqref{reach}, \eqref{r} and \eqref{s}} \}, \\
		\mathfrak{P}_{GPR} &= \{x \in \mathfrak{P} \mid x \ \text{satisfies \eqref{reach}, \eqref{r} and \eqref{g}} \},
	\end{align*}
	with the formulas
	\begin{align}
		\tag{reach}
		a^x_{s,t} = \!\!\!\sum_{\alpha \in \Act(s)}\!\!\! x_{s, \alpha} &\cdot \sum_{u \in S} P(s, \alpha, u) \cdot a^x_{u,t},\\
		\sum_{c \in \Cause} a^x_{\init, c} &> 0,
		\tag{r} \\
		\sum_{\eff \in \Eff}\!\!\! a^x_{c, \eff}  > \sum_{\eff \in \Eff}\!\!\! a^x_{\init, \eff} \ \ \ &\text{for all} \ c \in \Cause \ \text{with} \ a^x_{\init, c} > 0,
		\tag{s}\\\
		\tp^x \cdot \tn^x &- \fp^x \cdot \fn^x > 0.
		\tag{g}
	\end{align}
	Here, $tp^x, tn^x, fp^x$ and $fn^x$ are as defined by the confusion matrix (Figure \ref{fig:confusion}).
	
	\begin{ctheorem}{9}
		For an MDP $\cM$, a set of terminal states $\Eff$ and a predictor set $\Cause$ for which a memoryless SPR (resp. GPR) policy exists, the sets $\mathfrak{P}$ and $\mathfrak{P}_{SPR}$ (resp. $\mathfrak{P}_{GPR}$) have the same dimension.
	\end{ctheorem}
	\begin{proof}	
		If $\cM$ is a DTMC, then the dimensions of both $\mathfrak{P}$ and $\mathfrak{P}_{SPR}$ (and $\mathfrak{P}_{GPR}$) are $0$ since there is exactly one policy.
		So, we assume there is some state $s \in S$ with $\Act(s) > 1$.
		Moreover, we assume that $\Cause$ is not a universal SPR (resp. GPR) cause by \cite{LMCS24}, since otherwise $\mathfrak{P} = \mathfrak{P}_{SPR}$ (resp. $\mathfrak{P} = \mathfrak{P}_{GPR}$).
		In order to proof the claim we will show that for any vector $x \in \mathfrak{P}_{SPR}$ (resp. $\mathfrak{P}_{GPR}$) there is an $\varepsilon > 0$ such that the $\varepsilon$-ball at $x$ in $\mathfrak{P}$, given by $B^{\mathfrak{P}}_{x, \varepsilon} = \{y \in \mathfrak{P} \mid \lVert x - y \rVert < \varepsilon\}$, is completely contained in $\mathfrak{P}_{SPR}$ (resp. in $\mathfrak{P}_{GPR}$).
		The set $B^{\mathfrak{P}}_{x, \varepsilon}$ is the open $\varepsilon$-ball inside $\mathfrak{P}$ around $x$ which is the intersection of the real $\varepsilon$-ball $B_{x, \varepsilon} \subset \Real^\StAct$ with $\mathfrak{P}$.
		This way $B^{\mathfrak{P}}_{x, \varepsilon}$ has the same dimension as $\mathfrak{P}$.
		
		For $c \in \Cause$ we define $f_{c}: \mathfrak{P} \to \Real$ by
		\begin{align*}
			f_c(x) = 	\sum_{\eff \in \Eff} a^x_{c, \eff} - \sum_{\eff \in \Eff} a^x_{\init, \eff}.
		\end{align*}
		Then, $f_c$ is continuous for all $c \in \Cause$.
		Now, let $x \in \mathfrak{P}_{SPR}$.
		By the SPR condition we have $f_c(x) > 0$ for all $c \in \Cause$.
		Since $\mathfrak{P}$ is simply connected (as a polytope given by a linear program) and by the assumption that $\Cause$ is not a universal SPR cause, there are $c \in \Cause$ and $y \in \mathfrak{P}$ such that $f_c(y) = 0$.
		For this $c$ we can choose $y \in \mathfrak{P}$ in a way such that $\lVert x - y \rVert$ is minimal.
		Let $\varepsilon$ be this minimal distance $\varepsilon = \lVert x - y \rVert$.
		
		We now show that for all $z \in B^{\mathfrak{P}}_{x, \varepsilon}$ we have $f_c(z) > 0$ and thus $B^{\mathfrak{P}}_{x, \varepsilon} \subseteq \mathfrak{P}_{SPR}$.
		Assume $z \in B^{\mathfrak{P}}_{x, \varepsilon}$ with $f_c(z) \leq 0$.
		If $f_c(z) < 0$ then there is also $z' \in B^{\mathfrak{P}}_{x, \varepsilon}$ with $f_c(z') = 0$ by continuity of $f_c$ and the intermediate value theorem.
		Thus we assume w.l.o.g $f_c(z) = 0$ as otherwise we can consider $z'$.
		But then $f_c(z) = 0$ and $\lVert x - z \rVert < \varepsilon = \lVert x - y \rVert$ contradict the minimality of our choice of $y$.
		
		Since $B^{\mathfrak{P}}_{x, \varepsilon} \subseteq \mathfrak{P}_{SPR}$ and $B^{\mathfrak{P}}_{x, \varepsilon}$ has the same dimension as $\mathfrak{P}$, we conclude that $\mathfrak{P}_{SPR}$ also has the same dimension as $\mathfrak{P}$.
		Note that for the proof we only used the continuity of both sides of the SPR condition.
		Since the GPR condition is also continuous, we can apply the same arguments to show that also $\mathfrak{P}_{GPR}$ and $\mathfrak{P}$ have the same dimension.
	\end{proof}
	
	\subsection{Proofs of Section 4.2}
	
	We give a more detailed definition of the canonical MDP:
	
	\begin{cdefinition}{12}[Canonical MDP]
		We transform the original MDP $\cM$ to the \emph{canonical MDP} $\canMDP{\cM}{\Cause}$ in the following way:
		\begin{itemize}
			\item[(i)] All outgoing transitions from states $c \in \Cause$ are deleted and instead two fresh actions $\alpha_{\min}$ and $\alpha_{\max}$ are added.
			They transition to a new state $\TP$ with the minimal and resp. maximal probability to reach $\Eff$ from $c$.
			With the remaining probability they transition to a new state $\FP$, that is:
			\begin{align*}
				P'(c, \alpha_{\min}, \TP) & = p_{c,\min} \eqdef \Pr^{\min}_{\cM, c}(\lozenge \Eff), \\
				P'(c, \alpha_{\min}, \FP) &= 1 - p_{c,\min}, \\
				P'(c, \alpha_{\max}, \TP) &= p_{c,\max} \eqdef \Pr^{\max}_{\cM, c}(\lozenge \Eff),\\
				P'(c, \alpha_{\max}, \FP) &= 1 - p_{c,\max},
			\end{align*}
			where $\TP$ and $\FP$ are fresh terminal states.
			\item[(ii)] Collapse the maximal end-components (MECs) $\cE$ of the resulting MDP into single states $s_{\cE}$ by taking the MEC-quotient, see e.g. \cite{Alfaro-CONCUR99}.
			\item[(iii)] Collapse the states of $\Eff$ to a fresh state $\FN$ and other terminal states to the fresh state $\TN$.
			States $s_{\cE}$ representing MECs get the additional transition $P(s_{\cE} , \tau, \TN) = 1$.
			The resulting MDP is $\canMDP{\cM}{\Cause}$.
		\end{itemize}
		In $\canMDP{\cM}{\Cause}$ we consider $\Eff' = \{\TP, \FN\}$.
		%	If $\Cause = \{c\}$ is a singleton we write $\canMDP{\cM}{c}$.
		\Ende
	\end{cdefinition}
	
	For the soundness of the canonical MDP, two results are stated: Lemma \ref{lem:preserving confusion matrix} and Corollary \ref{cor:preserving PR}.
	Note that the construction of the canonical MDP is very similar to \cite{LMCS24}.
	However, in our version there are differences in the enabled actions in states $c \in \Cause$.
	We will provide proofs for both results here, but  point out that the techniques are very close to \cite{LMCS24}.
	In particular, the usage of the correspondence of an MEC-quotient to its original MDP is used analogously.
	
	\pagebreak
	
	\begin{clemma}{13}[$\canMDP{\cM}{\Cause}$ preserves Confusion Matrix]
		Given an MDP $\cM$, a set of terminal states $\Eff \subset S$ and a set $\Cause \subseteq S \setminus \Eff$, for each policy $\sched$ of $\cM$ there is a policy $\usched$ of $\canMDP{\cM}{\Cause}$ and vice versa such that
		\begin{center}
			$\tp^\sched_\cM = \Pr^\usched_{\canMDP{\cM}{\Cause}}(\lozenge \TP), \qquad \fp^\sched_\cM = \Pr^\usched_{\canMDP{\cM}{\Cause}}(\lozenge \FP),$ \\
			$\fn^\sched_\cM = \Pr^\usched_{\canMDP{\cM}{\Cause}}(\lozenge \FN), \qquad \tn^\sched_\cM = \Pr^\usched_{\canMDP{\cM}{\Cause}}(\lozenge \TN).$
		\end{center}
	\end{clemma}
	\begin{proof}
		We will proof the lemma by showing two claims:
		\begin{claim*}[i]
			For each policy $\sched$ of $\cM$ there is a policy $\tsched$ of the intermediate MDP $\minmaxMDP{\cM}{\Cause}$ [after step (i) of Def. 12] and vice versa such that
			\begin{align*}
				\tp^\sched_\cM &= \Pr^\tsched_{\minmaxMDP{\cM}{\Cause}}(\lozenge \TP) = \tp^\tsched_{\minmaxMDP{\cM}{\Cause}},\\
				\fp^\sched_\cM &= \Pr^\tsched_{\minmaxMDP{\cM}{\Cause}}(\lozenge \FP) = \fp^\tsched_{\minmaxMDP{\cM}{\Cause}},\\
				\fn^\sched_\cM &= \fn^\tsched_{\minmaxMDP{\cM}{\Cause}}, \\
				 \tn^\sched_\cM & = \tn^\tsched_{\minmaxMDP{\cM}{\Cause}},
			\end{align*}
		\end{claim*}
		``$\implies$'':
		Given a policy $\sched$ of $\cM$ we define a corresponding policy $\tsched$ of $\minmaxMDP{\cM}{\Cause}$ in the following way:
		For all paths $\pi \in \Paths_\fin(\minmaxMDP{\cM}{\Cause})$ ending in states $s \in S \setminus \Cause$ we define for all actions $\alpha \in \Act(s)$: $\tsched(\pi, \alpha) = \sched(\pi, \alpha)$.
		When a path $\pi \in \Paths_\fin(\minmaxMDP{\cM}{\Cause})$ reaches a state $c \in \Cause$ we let
		\begin{align*}
			\tsched(\pi, \alpha_{\max}) &= \frac{\Pr_{\cM, c}^{\residual{\sched}{\pi}}(\lozenge \Eff) - p_{c,\min}}{p_{c,\max} - p_{c,\min}} \quad \text{and} \\ \tsched(\pi, \alpha_{\min}) &= 1 - \tsched(\pi, \alpha_{\max}),
		\end{align*}
		if $p_{c,\max} > p_{c,\min}$.
		Otherwise, if $p_{c,\max} = p_{c,\min}$ then the actions $\alpha_{\max}$ and $\alpha_{\min}$ give equivalent distributions for the transition function and we let $\tsched(\pi, c_0) = \alpha_{\min}$.
		
		In order to now show that $\tp^\sched_\cM = \Pr^\tsched_{\minmaxMDP{\cM}{\Cause}}(\lozenge \TP)$ we need to show that for any path $\pi \in \Paths_\fin(\cM)$ that does not visit $\Cause$ or only visits $\Cause$ once with $\last(\pi) \in \Cause$ we have 
		\begin{align*}
			x := \Pr_\cM^{\residual{\sched}{\pi}}(\lozenge \Cause \wedge \lozenge \Eff) = \Pr_{\minmaxMDP{\cM}{\Cause}}^{\residual{\tsched}{\pi}}(\lozenge \TP) =: y,
		\end{align*}
		where $\pi$ is considered as a path in both MDPs $\cM$ and $\minmaxMDP{\cM}{\Cause}$ by abuse of notation.
		For paths not visiting a state $c \in \Cause$ this is clear by construction.
		If $\pi$ ends in a state $c \in \Cause$ we have the following
		\begin{align*}
			y & = \frac{x - p_{c,\min}}{p_{c,\max} - p_{c,\min}}p_{c,\max} + \left(1- \frac{x - p_{c,\min}}{p_{c,\max} - p_{c,\min}} \right) p_{c,\min} \\
			& = p_{c,\min} \! +\! \frac{p_{c,\max}\! \cdot\! x - p_{c,\max}\! \cdot\! p_{c,\min} - p_{c,\min}\! \cdot\! x \! + \! (p_{c,\min})^2}{p_{c,\max} - p_{c,\min}} \\
			& = p_{c,\min} + \frac{(p_{c,\max} - p_{c,\min})(x - p_{c,\min})}{p_{c,\max} - p_{c,\min}} = x.
		\end{align*}
		A analogue calculation shows $\fp^\sched_\cM = \Pr^\tsched_{\minmaxMDP{\cM}{\Cause}}(\lozenge \FP)$.
		The remaining equalities follow from the fact that no path that never visits $C$ in the original MDP $\cM$ is influenced by the construction of $\minmaxMDP{\cM}{\Cause}$.
		
		``$\impliedby$'':
		Let $\tsched$ be a policy of $\minmaxMDP{\cM}{\Cause}$.
		We define a corresponding policy $\sched$ of $\cM$ by mimicking $\tsched$ before a state $c \in \Cause$ was visited.
		Whenever a path visits a state $c \in \Cause$ the policy $\tsched$ chooses $\alpha_{\max}$ with probability $p \in [0,1]$ and $\alpha_{\min}$ with probability $(1-p)$.
		The corresponding policy $\sched$ then chooses a maximizing action $\alpha \in \Act(c)$ for $\lozenge \Eff$ with probability $p$ and a minimizing action $\beta \in \Act(c)$ for $\lozenge \Eff$ with probability $1-p$.
		Afterwards, $\sched$ realizes a maximizing MD policy for $\lozenge \Eff$ if a maximizing action $\alpha$ was taken and otherwise it realizes a minimizing MD policy for $\lozenge \Eff$.
		This transformation adds two additional memory cells to $\sched$.
		One memory cell keeps track of whether $\Cause$ was visited or not and one memory cell keeps track whether a maximizing or minimizing action was taken in the first visited $c \in \Cause$.
		Furthermore, $\sched$ uses randomization.
		The claimed equalities then follow by construction of $\sched$.
		
		\begin{claim*}[ii + iii]
			For each policy $\tsched$ of $\minmaxMDP{\cM}{\Cause}$ [(i) of Def. 12] there is a policy $\usched$ of $\canMDP{\cM}{\Cause}$ and vice versa such that
			\begin{align*}
				\tp^\tsched_{\minmaxMDP{\cM}{\Cause}} &= \Pr^\usched_{\canMDP{\cM}{\Cause}}(\lozenge \TP), & \fp^\tsched_{\minmaxMDP{\cM}{\Cause}} &= \Pr^\usched_{\canMDP{\cM}{\Cause}}(\lozenge \FP), \\
				\fn^\tsched_{\minmaxMDP{\cM}{\Cause}} &= \Pr^\usched_{\canMDP{\cM}{\Cause}}(\lozenge \FN), & \tn^\tsched_{\minmaxMDP{\cM}{\Cause}} &= \Pr^\usched_{\canMDP{\cM}{\Cause}}(\lozenge \TN).
			\end{align*}
		\end{claim*}
		
		``$\implies$'':
		Given a policy $\tsched$ of $\minmaxMDP{\cM}{\Cause}$ we define $\usched$ in the following way:
		For paths which do not visit a state $s_\cE$ representing an MEC the policy $\usched$ mimics $\tsched$.
		Whenever a path enters an MEC $\cE$ of $\minmaxMDP{\cM}{\Cause}$ the corresponding path $\pi$ in $\canMDP{\cM}{\Cause}$ ends in the state $s_\cE$.
		The policy $\tsched$ stays in the MEC with probability $p \in [0,1]$ and we define $\usched(\pi,\tau) = p$, where $\tau$ is the action leading to $\TN$ with probability $1$.
		With the remaining probability $1-p$ policy $\tsched$ chooses external actions with some distribution.
		According to the same distribution weighted with $1-p$ we let $\usched$ choose the same external actions.
		
		Now, the first two equalities $\tp^\tsched_{\minmaxMDP{\cM}{\Cause}}  = \Pr^\usched_{\canMDP{\cM}{\Cause}}(\lozenge \TP)$ and $\fp^\tsched_{\minmaxMDP{\cM}{\Cause}}  = \Pr^\usched_{\canMDP{\cM}{\Cause}}(\lozenge \FP)$ follow from results on the MEC-quotient \cite{deAlfaro1997,Alfaro-CONCUR99}.
		The same argumentation works for the equality $\fn^\tsched_{\minmaxMDP{\cM}{\Cause}} = \Pr^\usched_{\canMDP{\cM}{\Cause}}(\lozenge \FN)$ as here only terminal states of $\minmaxMDP{\cM}{\Cause}$ are collapsed into a fresh terminal state $\FN$.
		For the last equality $\tn^\tsched_{\minmaxMDP{\cM}{\Cause}} = \Pr^\usched_{\canMDP{\cM}{\Cause}}(\lozenge \TN)$ we note that this directly corresponds to either reaching any terminal state $t \notin \Eff$ in $\minmaxMDP{\cM}{\Cause}$ or staying in an end-component indefinitely.
		The probability for the first case is redistributed to $\TN$ by collapsing exactly those states.
		For the second case the constructed policy $\usched$ chooses $\tau$ in a state $s_\cE$ representing the end-component with compatible probability.
		As the MEC quotient does not influence the probabilities in other ways \cite{deAlfaro1997, Alfaro-CONCUR99} the claim follows.
		
		``$\impliedby$'':
		Given a policy $\usched$ of $\canMDP{\cM}{\Cause}$ a corresponding policy $\tsched$ for $\minmaxMDP{\cM}{\Cause}$ can be defined using the correspondence from of an MEC-quotient to its original MDP \cite{deAlfaro1997,Alfaro-CONCUR99} which directly implies the claimed equalities.
		
		The Lemma now follows from the combination of both claims.
	\end{proof}
	
	\begin{ccorollary}{14}[$\canMDP{\cM}{\Cause}$ preserves Probability-Raising]
		Given an MDP $\cM$, a set of terminal states $\Eff \subset S$ and a set $\Cause \subseteq S \setminus \Eff$, a policy $\sched$ of $\cM$ satisfies \condS (resp. \condG ) for $\Cause$ and $\Eff$ in $\cM$ iff the corresponding policy $\usched$ from Lemma \ref{lem:preserving confusion matrix} satisfies \condS (resp. \condG ) for $\Cause$ and $\{\TP, \FN\}$ in $\canMDP{\cM}{\Cause}$.
	\end{ccorollary}

	\begin{proof}
		For (GPR) this follows directly from Lemma \ref{lem:preserving confusion matrix} as we can express the (GPR) condition in terms of the confusion matrix, as shown later.
		Thus, we consider the case where $\sched$ is satisfying (SPR) for $\Cause$ and $\Eff$ in $\cM$.
		Its corresponding policy $\usched$ of $\canMDP{\cM}{\Cause}$ is defined as described in the proof of Lemma \ref{lem:preserving confusion matrix}.
		Since $\sched$ is satisfying SPR in $\cM$, we have $\Pr^\usched_{\canMDP{\cM}{\Cause}}(\lozenge \Cause) > 0$ trivially.
		In $\canMDP{\cM}{\Cause}$ any state $c \in \Cause$ can only be visited once.
		Thus, all paths visiting such a state $c$ satisfy $((\neg \Cause) \until c)$.
		Therefore, we have
		\begin{align*}
			\Pr^\usched_{\canMDP{\cM}{\Cause}}(\lozenge \Eff \mid \lozenge c) &= \Pr^\sched_\cM(\lozenge \Eff \mid (\neg \Cause) \until c) \\
			&> \Pr^\sched_\cM(\lozenge \Eff) \\ 
			&= \Pr^\usched_{\canMDP{\cM}{\Cause}}(\lozenge \{\TP, \FN\}),
		\end{align*}
		where the second equality follows from Lemma \ref{lem:preserving confusion matrix}.
		
		For the other direction, the reversal of the above argumentation yields the claim.
	\end{proof}
	
	Next we want to proof the Characterization of SPR policies (Lemma \ref{lem:char-SPR-policy}).	
	Recall that we assume the canonical MDP $\cM = \canMDP{\cM}{\Cause}$ and consider the maximal probability to reach $\TP$ among all cause states: 
	\begin{align*}
		p^\star = \max_{c \in \Cause} p_{c, \max}
	\end{align*}
	Then, for any $p \in (0, p^\star]$ we define the transformed MDP $\cM_p = (S, \Act \cup \{\delta\}, P', \init)$ by
	\begin{itemize}
		\item only action  $\delta$ is enabled in $c \in \Cause$ with $P'(c, \delta, \TP) = \max \{p, p_{c, \min}\}$ and $P'(c, \delta, \FP) = 1 - P'(c, \delta, \TP)$,
		\item for $s \in S \setminus \Cause, \ t \in S$ and $\alpha \in \Act$ we have $P'(s, \alpha, t) = P (s, \alpha, t)$.
	\end{itemize}
	So, $\cM_p$ behaves as $\cM$ except when a state $c \in \Cause$ is reached, in which case the single enabled action leads to $\TP$ with $p$ as long as $p > p_{c, \min}$.
	We abbreviate $\cM_{p^\star}$ by $\cM^\star$.
	By construction, any policy $\usched$ of $\cM_p$ corresponds to a policy $\sched$ of $\cM$.
	
	\begin{clemma}{15}[Characterization of SPR Policies]
		For an MDP $\cM$ with set of terminal states $\Eff \subset S$ and set of states $\Cause \subseteq S \setminus \Eff$ there is an SPR policy $\sched$ for $\Cause$ and $\Eff$ in $\cM$ iff $\Pr^{\min}_{\cM^\star}(\lozenge \Eff) < p^\star$.
	\end{clemma}
	\begin{proof}
		The equivalence is shown by showing that the following are equivalent:
		\begin{description}
			\item[(1)] There is an SPR policy $\sched$ for $\Cause$ and $\Eff$ in $\cM$.
			\item[(2)] There exists a policy $\sched$ for $\cM$ such that
			\begin{itemize}
				\item $\Pr^\sched_\cM(\lozenge \Eff) < p^\star$ and
				\item if $c \in \Cause$ with $\Pr^\sched((\neg \Cause) \until c) > 0$ then (SPR) holds for $c$ under $\sched$.
			\end{itemize}
			\item[(3)] There exists $p \in (0,1]$ with $p \leq p^\star$ such that $\Pr^{\min}_{\cM_p}(\lozenge \Eff) < p$.
			\item[(4)] $\Pr^{\min}_{\cM^\star}(\lozenge \Eff) < p^\star$.
		\end{description}
		Obviously, (1) $\implies$ (2).
		For (2) $\implies$ (1) consider the following MR policy $\usched$ of $\cM$:
		\begin{itemize}
			\item For $s \in S \setminus \Cause$ define $\usched(s, \cdot)$ to be a uniform distribution over $\Act(s)$.
			\item For $c \in \Cause$ define $\usched(c, \alpha_{\max}) = 1$.
		\end{itemize}
		By definition, $\cM$ and $\cM_\usched$ have the same underlying graph.
		Let $\sched$ be the SPR policy for $\Cause$ and $\Eff$ in $\cM$ from (2).
		Since $\cM$ has no end-components, we can assume $\sched$ to be memoryless.
		Moreover, for a value $\varepsilon \in (0,1]$ we can consider the policy $\sched_\varepsilon = \varepsilon \usched \oplus (1 - \varepsilon) \sched$ in which we have for each state-action pair $(s, \alpha)$:
		\begin{align*}
			\sched_\varepsilon(s, \alpha) = \varepsilon \cdot \usched(s, \alpha) + (1 - \varepsilon) \cdot \sched(s, \alpha).
		\end{align*}
		For all states $s \in S$ we then get $\Pr^{\sched_\varepsilon}_\cM(\lozenge s) > 0$.
		Moreover, let
		\begin{align*}
			p_{\Cause}^\sched = \min_{c \in \Cause} \Pr^\sched_\cM(\lozenge \Eff \mid (\neg \Cause) \until c).
		\end{align*}
		By condition (2) we have $p_\Cause^\sched > \Pr^\sched_\cM(\lozenge \Eff)$ and we can consider $\varepsilon > 0$ with
		\begin{align*}
			\frac{\varepsilon}{1 - \varepsilon} < p_\Cause^\sched - \Pr^\sched_\cM(\lozenge \Eff).
		\end{align*}
		For such $\varepsilon$ we have
		\begin{align*}
			\Pr^{\sched_\varepsilon}_\cM(\lozenge \Eff) & \leq (1 - \varepsilon) \cdot \Pr^\sched_\cM(\lozenge \Eff) + \varepsilon \cdot \Pr^\usched_\cM(\lozenge \Eff) \\[0.5ex]
			& < (1 - \varepsilon) \cdot p_{\Cause}^\sched \\[0.5ex]
			& \leq (1 - \varepsilon) \cdot \Pr^\sched_\cM(\lozenge \Eff \mid (\neg \Cause) \until c) \\[0.5ex]
			& < \Pr^\sched_\cM(\lozenge \Eff \mid (\neg \Cause) \until c) \\[0.5ex]
			& \leq \Pr^{\sched_\varepsilon}_\cM(\lozenge \Eff \mid (\neg \Cause) \until c),
		\end{align*}
		which shows that $\sched_\varepsilon$ is an SPR policy for $\Cause$ and $\Eff$ in $\cM$.
		
		We next show (1) $\implies$ (3).
		Here, we are given the SPR policy $\sched$ for $\Cause$ and $\Eff$ in $\cM$.
		For 
		\begin{align*}
			p = \min_{c \in \Cause} \Pr^\sched_\cM(\lozenge \Eff \mid (\neg \Cause)),
		\end{align*}
		consider the policy $\tsched$ of $\cM_p$ that behaves identically to $\sched$ for all paths ending in non-terminal states $s \in S \setminus \Cause$.
		Furthermore, for $c \in \Cause$ consider the following sets of paths:
		\begin{align*}
			\Pi_c & = \{\pi = \init \ \cdots \ c \ \TP \mid \pi \ \text{is} \ \sched-\text{path}\}, \\
			\Pi_\FN & = \{\pi = \init \ \cdots \ \FN \mid \pi \ \text{is} \ \sched-\text{path}\} \quad \text{and} \\
			\Pi_\TP & = \bigcup_{c \in \Cause} \Pi_c.
		\end{align*}
		This allows us to express the reachability probabilities under $\sched$ and $\tsched$ by
		\begin{align*}
			\Pr^\sched_\cM(\lozenge \Eff) &= \sum_{\pi \in \Pi_\FN} \Pr^\sched_\cM(\pi) + \sum_{\pi \in \Pi_\TP} \Pr^\sched_\cM(\pi), \\[1ex]
			\Pr^\tsched_{\cM_p}(\lozenge \Eff) &= \sum_{\pi \in \Pi_\FN} \Pr^\sched_\cM(\pi) + \sum_{\pi \in \Pi_\TP} \Pr^\tsched_{\cM_p}(\pi).
		\end{align*}
		However, by choice of $p$ and definition of $\cM_p$ we have $\Pr^\sched_\cM(\pi) \geq \Pr^\tsched_{\cM_p}(\pi)$ for every $\pi \in \Pi_\TP$ which yields:
		\begin{align*}
			\Pr^{\min}_{\cM_p}(\lozenge \Eff) \leq \Pr^\tsched_{\cM_p}(\lozenge \Eff) \leq \Pr^\sched_\cM(\lozenge \Eff) < p.
		\end{align*}
		
		We now turn to (3) $\implies$ (4).
		So, let $p \in (0, p^\star)$ and $\Pr^{\min}_{\cM_p}(\lozenge \Eff) < p$.
		For $\cM^\star = \cM_{p^\star}$ we need to show $\Pr^{\min}_{\cM^\star}(\lozenge \Eff) < p^\star$.
		Let $\tsched$ be a policy of $\cM_p$ which minimizes the probability of reaching $\Eff$ in $\cM_p$.
		That is, $\Pr^\tsched_{\cM_p}(\lozenge \Eff) = \Pr^{\min}_{\cM_p}(\lozenge \Eff) < p$.
		By using the abbreviation $p_c^{\min} = \Pr^{\min}_\cM(\lozenge \Eff \mid (\neg \Cause) \until c)$, we proceed to partition $\Cause$ into the sets
		\begin{align*}
			\Cause_{\leq} &= \{c \in \Cause \mid p_c^{\min} \leq p\}, \\
			\Cause_{\operatorname{in}} &= \{c \in \Cause \mid p <p_c^{\min} \leq p^\star\}, \\
			\Cause_{>} &= \{c \in \Cause \mid p_c^{\min} > p^\star\}.
		\end{align*}
		We also use the following abbreviations:
		\begin{align*}
			\varrho & = \Pr^\tsched_{\cM_p}(\lozenge \Cause_{\leq}), \ \text{for} \ c \in \Cause_{\operatorname{in}}: \ \varrho_c = \Pr^\tsched_{\cM_p}(\lozenge c), \\
			R & = \sum_{c \in \Cause_{>}} \Pr^\tsched_{\cM_p}(\lozenge c) \cdot p_c^{\min} + \Pr^\tsched_{\cM_p}(\lozenge \FN).
		\end{align*}
		We can thus express the effect probability in $\cM_p$ under $\tsched$ by
		\begin{align*}
			\Pr^\tsched_{\cM_p}(\lozenge \Eff) = \varrho \cdot p + \sum_{c \in \Cause_{\operatorname{in}}} \varrho_c \cdot p_c^{\min} + R 
		\end{align*}
		When viewing $\tsched$ as a policy for $\cM^\star$ we get
		\begin{align*}
			\Pr^\tsched_{\cM_p}(\lozenge \Eff) = \varrho \cdot p^\star + \sum_{c \in \Cause_{\operatorname{in}}} \varrho_c \cdot p^\star + R = r \cdot p^\star + R,
		\end{align*}
		where $r = \varrho + \sum_{c \in \Cause_{\operatorname{in}}} \varrho_c$.
		We use $0 \leq p^\star {-} p_c^{\min} < p^\star {-} p$ to get
		\begin{align*}
			\Pr^\tsched_{\cM^\star}(\lozenge \Eff) {-} \Pr^\tsched_{\cM_p}(\lozenge \Eff) &= \varrho (p^\star {-} p) + \!\!\!\!\!\sum_{c \in \Cause_{\operatorname{in}}}\!\!\!\!\! \varrho_c (p^\star {-} p_c^{\min}) \\
			&\leq r (p^\star {-} p).
		\end{align*}
		We obtain:
		\begin{align*}
			\Pr^{\min}_{\cM^\star}(\lozenge \Eff) & \leq \Pr^\tsched_{\cM^\star} \leq \Pr^\tsched_{\cM_p}(\lozenge \Eff) + r(p^\star - p) \\
			& \leq \Pr^\tsched_{\cM_p}(\lozenge \Eff) + p^\star - p \\
			& = p^\star (p - \Pr^\tsched_{\cM_p}(\lozenge \Eff)) < p^\star,
		\end{align*}
		where we use $r \leq 1$ and $\Pr^\tsched_{\cM_p}(\lozenge \Eff) < p$.
		
		We are left to proof (4) $\implies$ (2).
		Here, we consider an MD policy $\tsched$ of $\cM^\star$ which realizes the equality $\Pr^\tsched_{\cM^\star}(\lozenge \Eff) = \Pr^{\min}_{\cM^\star}(\lozenge \Eff)$.
		The corresponding policy $\sched$ of $\cM$ to $\tsched$ is a witness for condition (2).
	\end{proof}
	
	\paragraph{Reformulating GPR}
	Here, we are proving the inequality describing (GPR) used in Section 4.2.
	From this inequality the formula (freq-GPR) is derived in the paper.
	
	\begin{clemma}{}[Alternative GPR Condition]
		\label{lem:GPR-poly-constraint}
		In $\cM = \canMDP{\cM}{\Cause}$ the set $\Cause$ satisfies \condG if and only if for each policy $\sched$ with $\Pr^{\sched}(\lozenge \Cause) > 0$ the following condition holds:
		\begin{align*}
			\label{GPR-1}
			0 > \Pr^\sched(\lozenge \FN) \cdot \Pr^\sched(\lozenge \FP) - \Pr^\sched(\lozenge \TN) \cdot \Pr^\sched(\lozenge \TP)
			\tag{GPR-1}
		\end{align*}
	\end{clemma}               
	
	\begin{proof}
		We fix a policy $\sched$ of $\cM$ with $\Pr^\sched(\lozenge \Cause) > 0$ and use the following abbreviations
		\begin{align*}
			\Pr^\sched(\lozenge \TP) = tp, && \Pr^\sched(\lozenge \FP) = fp, \\
			\Pr^\sched(\lozenge \FN) = fn, && \Pr^\sched(\lozenge \TN) = tn.
		\end{align*}
		By construction, the four states $\TP, \FP, \FN, \TN$ are the only terminal states in the MEC-free MDP $\cM$ and thus we have
		\begin{align}
			\label{eq:four-terminals}
			1 = tp + fp + fn + tn.
			\tag{1}
		\end{align}
		Now (GPR) is equivalent to \eqref{GPR-1} by the following:
		\begin{align*}
			\text{(GPR)} &\Longleftrightarrow \Pr^\sched(\lozenge \Eff \mid \lozenge \Cause) > \Pr^\sched(\lozenge \Eff) \\[0.5ex]
			&\Longleftrightarrow \frac{\Pr^\sched(\lozenge \Eff \wedge \lozenge \Cause)}{\Pr^\sched(\lozenge \Cause)} > \Pr^\sched(\lozenge \Eff) \\[0.5ex]
			&\Longleftrightarrow \frac{tp}{tp+fp} > tp + fn \\[0.5ex]
			&\Longleftrightarrow 0 > tp + fn - \frac{tp}{tp+fp} \\[0.5ex]
			&\Longleftrightarrow 0 > tp^2 + tp \cdot fp + fn \cdot tp + fn \cdot fp - tp \\[0.5ex]
			&\Longleftrightarrow 0 > fn \cdot fp + tp \cdot (tp + fp + fn - 1) \\[0.5ex]
			&\stackrel{\textrm{\tiny \eqref{eq:four-terminals}}}{\Longleftrightarrow} 0 >  fn \cdot fp - tn \cdot tp \\[0.5ex]
			&\Longleftrightarrow 0 > \Pr^\sched\!(\!\lozenge \FN) \Pr^\sched\!(\!\lozenge \FP)\! - \! \Pr^\sched\!(\!\lozenge \TN) \Pr^\sched\!(\!\lozenge \TP) \\[0.5ex] &\Longleftrightarrow \text{\eqref{GPR-1}}
		\end{align*}
	\end{proof}

	\begin{ctheorem}{19}[Complexity GPR Policies Check]
		Deciding whether there is a GPR policy for $\Cause$ and $\Eff$ in $\cM$ can be done in $\NP$.
	\end{ctheorem}
	\begin{proof}
		By Lemma \ref{lem:preserving confusion matrix} and Corollary \ref{cor:preserving PR} we will assume $\cM = \canMDP{\cM}{\Cause}$ throughout the proof.
		Define the polytope $P$ of feasible state-action frequencies for a policy $\sched$ of $\cM$ defined by inequalities (S1)-(S3) by
		\begin{align*}
			P = \{x \in \Real^{\StAct} \mid x \ \text{satisfies (S1) - (S3)}\}
		\end{align*}
		and the function $f: P \to \Real$ with
		\begin{align*}
			f(x) = x_\TP \cdot x_\TN - x_\FP \cdot x_\FN.
		\end{align*}
		By definition $P$ is an $m = \lvert \StAct \rvert - \lvert S \rvert$ dimensional subspace of $\Real^{\lvert \StAct \rvert}$ since each equation given by (S2) and (S3) takes away one dimension.
		For the remaining proof, however, we will consider $P$ as a subset of $\Real^m$.
		We note that $f$ is continuous as it is quadratic and that $P$ is closed and convex as it is defined solely by linear equations and inequalities.
		If there are both $x_0 \in P$ with $f(x_0) < 0$ and $x_1 \in P$ with $f(x_1) > 0$ then there is also $x \in P$ with $f(x) = 0$ by the intermediate value theorem.
		We thus get three disjoint cases
		\begin{description}
			\item[Case 1.] For all $x \in P$ we have $f(x) > 0$.
			\item[Case 2.] For all $x \in P$ we have $f(x) < 0$.
			\item[Case 3.] There is $x \in P$ with $f(x) = 0$.
		\end{description}
		In order to decide in which case we are, we define the set $P_0$ of zero points of $f$ in $P$, that is, $P_0 = \{x \in P \mid f(x) = 0\}$.
		Deciding whether this set is empty can be done by quadratic programming in $\NP$ by \cite{Vavasis1990}.
		If the set is empty then we take any point $x \in P$ and evaluate $f(x)$ in polynomial time to decide whether we are in case 1 or case 2.
		In case 1 all MR policies $\sched$ of $\cM$ satisfy (GPR) and thus $\Cause$ is even a universal GPR cause by \cite{LMCS24}.
		In case 2 there is no MR policy satisfying (GPR) and thus no GPR policy.
		If the set $P_0$ is non-empty we have to do further analysis.
		
		So we consider case 3.
		Recall that the \emph{interior} $\overset{\circ}{S}$ of a set $S \subseteq \Real^m$ is the set of all $x \in S$ for which there is an $\varepsilon > 0$ such that $\{y \in \Real^m \mid \lVert y - x \rVert \leq \varepsilon\} \subseteq S$.
		Furthermore, if $S$ is a closed set then its \emph{boundary} is $\partial S = S \setminus \overset{\circ}{S}$.
		Since $P_0$ is non-empty and closed there are two further disjoint cases
		\begin{description}
			\item[Case 3.1] All $x \in P_0$ are on the boundary $\partial P$ of $P$.
			\item[Case 3.2] There is some $x \in P_0$ that is in the interior $\overset{\circ}{P}$ of $P$.
		\end{description}
		If we are in case 3.2 we can non-deterministically guess a point $x \in P_0$ and confirm that $x \notin \partial P$ in polynomial time, since $\partial P$ can be computed in polynomial-time as the boundary of a convex polytope given by a linear program.
		Note, that the encoding of $x \in P_0$ is also bounded by \cite{Vavasis1990}  as it is a solution of a quadratic program.
		Therefore, we can decide whether we are in case 3.2 in $\NP$ if not, then we are in case 3.1.
		
		In case 3.1 either all interior points $x \in \overset{\circ}{P}$ of $P$ satisfy $f(x) > 0$ or all such points satisfy $f(x) < 0$ as otherwise there would be a point $x \in \overset{\circ}{P}$ with $f(x) = 0$ by the intermediate value theorem as $P$ is convex and $f$ is continuous.
		Therefore, we can take any point $x \in \overset{\circ}{P}$ and check whether $f(x) > 0$.
		If yes, then there is a policy $\sched$ corresponding to $x$ which is also a GPR policy for $\Cause$ and $\Eff$ in $\cM$.
		If not, then there is no GPR policy for $\Cause$ and $\Eff$ in $\cM$.
		
		We are left with case 3.2.
		So, let $x \in P_0 \cap \overset{\circ}{P}$.
		Intuitively, we will move from $x$ in the direction of $\nabla f(x)$ while staying inside of $P$.
		This way, we arrive at a point $x^\prime \in P$ for which $f(x^\prime) < 0$.
		Note, that there are no outgoing actions from the states $\TP, \TN, \FP$ and $\FN$.
		Thus, the values $x_\TP, x_\TN, x_\FP$ and $x_\FN$ directly appear in $x$, w.l.o.g these are the last four coordinates of $x$ in the order as listed.
		Moreover, since they are the only four terminal states and $\cM_{\Cause}$ has no end-components we also have $x_\TP + x_\TN + x_\FP + x_\FN = 1$.
		We now consider the gradient $\nabla f(x)$ of $f$ at $x$.
		Recall, that the gradient of a function $f$ is representing the direction of the maximal increase of $f$ and its norm is describing the rate of the maximal increase.
		Since $f$ only uses the values $x_\TP, x_\TN, x_\FP$ and $x_\FN$, we can ignore all other coordinates for the notation of $\nabla f(x)$.
		So for $\nabla f(x)$ we we have
		\begin{align*}
			\nabla f(x) = \begin{bmatrix} x_\TN \\ x_\TP \\ - x_\FN \\ - x_\FP \end{bmatrix} \text{and} \ \lVert \nabla f(x) \rVert = \sqrt{x_\TN^2 \! + \! x_\TP^2 \! + \! x_\FN^2 \! + \! x_\FP^2}.
		\end{align*}
		Since $x_\TP + x_\TN + x_\FP + x_\FN = 1$ we also get $\lVert \nabla f(x) \rVert > 0$ for all $x \in P$.
		Thus, for all $x \in P$ the function $f$ is always increasing in direction $\nabla f(x)$ and decreasing in direction $- \nabla f(x)$.
		Futhermore, it has no local extreme points inside of $P$.
		
		Recall, that we consider case 3.2 and $x \in P_0 \cap \overset{\circ}{P}$.
		Since $x$ is in the interior of $P$, there exists an $\varepsilon > 0$ such that $x^\prime = x + \varepsilon \cdot \nabla f(x) \in P$.
		As the value of $f(x)$ is increasing in direction $\nabla f(x)$ we then get $f(x^\prime) > 0$.
		Since additionally $x^\prime \in P$ is indeed a solution to (S1)-(S3) and (freq-GPR), an MR policy $\sched$ which corresponds to $x^\prime$ constitutes a GPR policy for $\Cause$ and $\Eff$  in $\cM$.
	\end{proof}

\bibliography{main.bib}

\begin{thebibliography}{37}
\providecommand{\natexlab}[1]{#1}

\bibitem[{Baier, Funke, and Majumdar(2021)}]{IJCAI21}
Baier, C.; Funke, F.; and Majumdar, R. 2021.
\newblock A Game-Theoretic Account of Responsibility Allocation.
\newblock In Zhou, Z., ed., \emph{30th International Joint Conference on
  Artificial Intelligence (IJCAI)}, 1773--1779. ijcai.org.

\bibitem[{Baier and Katoen(2008)}]{BaierK2008}
Baier, C.; and Katoen, J.-P. 2008.
\newblock \emph{Principles of Model Checking (Representation and Mind Series)}.
\newblock The MIT Press, Cambridge, MA.
\newblock ISBN 026202649X, 9780262026499.

\bibitem[{Baier, Piribauer, and Ziemek(2024)}]{LMCS24}
Baier, C.; Piribauer, J.; and Ziemek, R. 2024.
\newblock Foundations of probability-raising causality in Markov decision
  processes.
\newblock \emph{Logical Methods in Computer Science}, Volume 20, Issue 1.

\bibitem[{Baier et~al.(2024)Baier, van~den Bossche, Klüppelholz, Lehmann, and
  Piribauer}]{AAAI24}
Baier, C.; van~den Bossche, R.; Klüppelholz, S.; Lehmann, J.; and Piribauer,
  J. 2024.
\newblock Backward Responsibility in Transition Systems Using General Power
  Indices.
\newblock \emph{Proceedings of the AAAI Conference on Artificial Intelligence},
  38(18): 20320--20327.

\bibitem[{Baldoni et~al.(2011)Baldoni, Berline, de~Loera, K{\"o}ppe, and
  Vergne}]{BBdLKV-integrating-over-simplex-2011}
Baldoni, V.; Berline, N.; de~Loera, J.~A.; K{\"o}ppe, M.; and Vergne, M. 2011.
\newblock {How to Integrate a Polynomial over a Simplex}.
\newblock \emph{{Mathematics of Computation}}, 80(273): 297--325.

\bibitem[{Beckers, Chockler, and Halpern(2023)}]{BCH23}
Beckers, S.; Chockler, H.; and Halpern, J.~Y. 2023.
\newblock A Causal Analysis of Harm.
\newblock arXiv:2210.05327.

\bibitem[{Braham and van Hees(2012)}]{BrahamvanHees2012}
Braham, M.; and van Hees, M. 2012.
\newblock An Anatomy of Moral Responsibility.
\newblock \emph{Mind}, 121 (483): 601--634.

\bibitem[{Brightwell and Winkler(1991)}]{BW1991}
Brightwell, G.; and Winkler, P. 1991.
\newblock Counting Linear Extensions.
\newblock \emph{Order}, 8: 225--242.

\bibitem[{Bárány and Füredi(1987)}]{Barany1987}
Bárány, I.; and Füredi, Z. 1987.
\newblock Computing the Volume is Difficult.
\newblock \emph{Discrete \& Computational Geometry}, 2: 319--326.

\bibitem[{Chadha, Sistla, and Viswanathan(2009)}]{ChadhaSV09}
Chadha, R.; Sistla, A.~P.; and Viswanathan, M. 2009.
\newblock On the Expressiveness and Complexity of Randomization in Finite State
  Monitors.
\newblock \emph{J. ACM}, 56(5).

\bibitem[{Chicco and Jurman(2020)}]{chicco2020advantages}
Chicco, D.; and Jurman, G. 2020.
\newblock The advantages of the {M}atthews correlation coefficient ({MCC}) over
  {F1} score and accuracy in binary classification evaluation.
\newblock \emph{BMC genomics}, 21(1): 1--13.

\bibitem[{Chockler and Halpern(2004)}]{ChocklerH04}
Chockler, H.; and Halpern, J.~Y. 2004.
\newblock Responsibility and Blame: A Structural-Model Approach.
\newblock \emph{J. Artif. Int. Res.}, 22(1): 93–115.

\bibitem[{Chockler, Halpern, and Kupferman(2008)}]{ChocklerHK2008}
Chockler, H.; Halpern, J.~Y.; and Kupferman, O. 2008.
\newblock What causes a system to satisfy a specification?
\newblock \emph{{ACM} Transactions on Computational Logic}, 9(3): 20:1--20:26.

\bibitem[{Ciesinski et~al.(2008)Ciesinski, Baier, Größer, and Klein}]{QEST08}
Ciesinski, F.; Baier, C.; Größer, M.; and Klein, J. 2008.
\newblock Reduction Techniques for Model Checking Markov Decision Processes.
\newblock In \emph{2008 Fifth International Conference on Quantitative
  Evaluation of Systems}, 45--54.

\bibitem[{Clarke, Grumberg, and Peled(1999)}]{CGP99}
Clarke, E.~M.; Grumberg, O.; and Peled, D. 1999.
\newblock \emph{Model Checking}.
\newblock MIT Press.

\bibitem[{de~Alfaro(1997)}]{deAlfaro1997}
de~Alfaro, L. 1997.
\newblock \emph{Formal Verification of Probabilistic Systems}.
\newblock Phd thesis, Stanford University, Stanford, USA.

\bibitem[{de~Alfaro(1999)}]{Alfaro-CONCUR99}
de~Alfaro, L. 1999.
\newblock Computing Minimum and Maximum Reachability Times in Probabilistic
  Systems.
\newblock In Baeten, J. C.~M.; and Mauw, S., eds., \emph{10th International
  Conference on Concurrency Theory (CONCUR)}, volume 1664 of \emph{Lecture
  Notes in Computer Science}, 66--81. Springer.

\bibitem[{{De Loera} et~al.(2013){De Loera}, Dutra, Köppe, Moreinis, Pinto,
  and Wu}]{deLoera2013}
{De Loera}, J.; Dutra, B.; Köppe, M.; Moreinis, S.; Pinto, G.; and Wu, J.
  2013.
\newblock Software for exact integration of polynomials over polyhedra.
\newblock \emph{Computational Geometry}, 46(3): 232--252.

\bibitem[{Duriqi et~al.(2024)Duriqi, Snop{\c{c}}e, Salihu, and
  Luma}]{DBLP:conf/meco/DuriqiSSL24}
Duriqi, B.; Snop{\c{c}}e, H.; Salihu, A.; and Luma, A. 2024.
\newblock An overview of parallel processing of rectangular determinant
  calculation.
\newblock In \emph{13th Mediterranean Conference on Embedded Computing, {MECO}
  2024, Budva, Montenegro, June 11-14, 2024}, 1--7. {IEEE}.

\bibitem[{Hall(2004)}]{Hall2004}
Hall, N. 2004.
\newblock Two Concepts of Causation.
\newblock In Collins, J.; Hall, N.; and Paul, L., eds., \emph{Causation and
  Counterfactuals}, 225--276. MIT Press.

\bibitem[{Halpern(2015)}]{Halpern15}
Halpern, J.~Y. 2015.
\newblock A Modification of the {H}alpern-{P}earl Definition of Causality.
\newblock In \emph{Proceedings of IJCAI'15}, 3022–3033. AAAI Press.
\newblock ISBN 9781577357384.

\bibitem[{Halpern and Kleiman-Weiner(2018)}]{Halpern2018TowardsFD}
Halpern, J.~Y.; and Kleiman-Weiner, M. 2018.
\newblock Towards Formal Definitions of Blameworthiness, Intention, and Moral
  Responsibility.
\newblock In \emph{AAAI Conference on Artificial Intelligence}.

\bibitem[{Halpern and Pearl(2005)}]{HalpernP04}
Halpern, J.~Y.; and Pearl, J. 2005.
\newblock {Causes and Explanations: A Structural-Model Approach. Part I:
  Causes}.
\newblock \emph{The British Journal for the Philosophy of Science}, 56(4):
  843--887.

\bibitem[{Junges, Torfah, and Seshia(2021)}]{JTS-RuntimeMonitorsMDP2021}
Junges, S.; Torfah, H.; and Seshia, S.~A. 2021.
\newblock Runtime Monitors for Markov Decision Processes.
\newblock In Silva, A.; and Leino, K. R.~M., eds., \emph{Computer Aided
  Verification}, 553--576. Cham: Springer International Publishing.
\newblock ISBN 978-3-030-81688-9.

\bibitem[{Kallenberg(2020)}]{Kallenberg20}
Kallenberg, L. 2020.
\newblock \emph{Lecture Notes {M}arkov Decision Problems - version 2020}.

\bibitem[{Kaltofen and Villard(2005)}]{DBLP:journals/cc/KaltofenV05}
Kaltofen, E.~L.; and Villard, G. 2005.
\newblock On the complexity of computing determinants.
\newblock \emph{Comput. Complex.}, 13(3-4): 91--130.

\bibitem[{Kozine and Utkin(2002)}]{KozineU02}
Kozine, I.; and Utkin, L.~V. 2002.
\newblock Interval-Valued Finite Markov Chains.
\newblock \emph{Reliab. Comput.}, 8(2): 97--113.

\bibitem[{Manna and Pnueli(1995)}]{MaPn95}
Manna, Z.; and Pnueli, A. 1995.
\newblock \emph{The Temporal Logic of Reactive and Concurrent Systems: Safety}.
\newblock Springer-Verlag.

\bibitem[{Mascle et~al.(2021)Mascle, Baier, Funke, Jantsch, and
  Kiefer}]{MBFJK21-Importance}
Mascle, C.; Baier, C.; Funke, F.; Jantsch, S.; and Kiefer, S. 2021.
\newblock Responsibility and verification: Importance value in temporal logics.
\newblock In \emph{2021 36th Annual ACM/IEEE Symposium on Logic in Computer
  Science (LICS)}, 1--14. Los Alamitos, CA, USA: IEEE Computer Society.

\bibitem[{Namjoshi(2001)}]{Namjoshi01}
Namjoshi, K.~S. 2001.
\newblock Certifying Model Checkers.
\newblock In \emph{13th International Conference on Computer Aided Verification
  (CAV)}, volume 2102 of \emph{Lecture Notes in Computer Science}, 2--13.
  Springer.

\bibitem[{Pearl(2009)}]{Pearl09}
Pearl, J. 2009.
\newblock \emph{Causality}.
\newblock Cambridge University Press, 2nd edition.

\bibitem[{Powers(2011)}]{Powers-fscore}
Powers, D. 2011.
\newblock Evaluation: From Precision, Recall and F-Measure to {ROC},
  Informedness, Markedness \& Correlation.
\newblock \emph{Journal of Machine Learning Technologies}, 2(1): 37--63.

\bibitem[{Press et~al.(2007)Press, Teukolsky, Vetterling, and
  Flannery}]{NumRecipies2007}
Press, W.~H.; Teukolsky, S.~A.; Vetterling, W.~T.; and Flannery, B.~P. 2007.
\newblock \emph{Numerical Recipes 3rd Edition: The Art of Scientific
  Computing}.
\newblock USA: Cambridge University Press, 3 edition.
\newblock ISBN 0521880688.

\bibitem[{Sen, Viswanathan, and Agha(2006)}]{SenVA-IntervalMC06}
Sen, K.; Viswanathan, M.; and Agha, G. 2006.
\newblock Model-Checking Markov Chains in the Presence of Uncertainties.
\newblock In Hermanns, H.; and Palsberg, J., eds., \emph{Tools and Algorithms
  for the Construction and Analysis of Systems}, 394--410. Berlin, Heidelberg:
  Springer Berlin Heidelberg.
\newblock ISBN 978-3-540-33057-8.

\bibitem[{Stein(1966)}]{Simplex-Volume}
Stein, P. 1966.
\newblock A Note on the Volume of a Simplex.
\newblock \emph{The American Mathematical Monthly}, 73(3): 299--301.

\bibitem[{Vavasis(1990)}]{Vavasis1990}
Vavasis, S.~A. 1990.
\newblock Quadratic programming is in {NP}.
\newblock \emph{Information Processing Letters}, 36(2): 73--77.

\bibitem[{Ziemek et~al.(2022)Ziemek, Piribauer, Funke, Jantsch, and
  Baier}]{ISSE22}
Ziemek, R.; Piribauer, J.; Funke, F.; Jantsch, S.; and Baier, C. 2022.
\newblock Probabilistic causes in {M}arkov chains.
\newblock \emph{Innovations in Systems and Software Engineering}.

\end{thebibliography}

\end{document}